\documentclass[%
 reprint,
superscriptaddress,
 amsmath,amssymb,
 aps,
pre,
]{revtex4-2}

\usepackage{amsmath,amssymb,amsfonts,amsthm}
\usepackage{graphicx}
\usepackage{bm,comment}
\usepackage{xcolor}
\usepackage{bbold}
\usepackage[normalem]{ulem}
\usepackage[colorlinks=true,allcolors=blue]{hyperref}
\usepackage[percent]{overpic}
\usepackage{hyperref,amsthm}

\usepackage{graphicx}
\usepackage{dcolumn}
\usepackage{bm}

\begin{document}

\preprint{APS/123-QED}

\title{Wave number dynamics in the real and complex Ginzburg-Landau equations}

\author{Yijun Lin}\email{yijunlinphys@mail.ustc.edu.cn}
 \affiliation{Department of Physics, University of California at Berkeley, Berkeley, CA 94720, USA}
\affiliation{Department of Physics, University of Science and Technology of China, Hefei 230026, China}
\author{Adrian van Kan}%
 \email{avankan@uni-muenster.de}
 \affiliation{Institut für Geophysik, Universität Münster, Corrensstr. 24, 48149 Münster, Germany}
\affiliation{Department of Mathematics, Texas A\&M University, College Station, TX 77843, USA}
\affiliation{Department of Physics, University of California at Berkeley, Berkeley, CA 94720, USA} 

 \author{Edgar Knobloch}
 \email{knobloch@berkeley.edu}
 \affiliation{Department of Physics, University of California at Berkeley, Berkeley, CA 94720, USA}

\date{\today}

\begin{abstract}
We investigate the large-scale dynamics of the one-dimensional real and complex Ginzburg–Landau equations using a scalar equation for the local wave number derived from a Wentzel–Kramers–Brillouin (WKB) expansion. This reduction captures a broad range of nonlinear phenomena. For the real Ginzburg-Landau equation, the reduced dynamics takes a conserved gradient form. After introducing a higher-order regularization, we obtain families of exact stationary solutions and show that all nonuniform steady states are linearly unstable. In particular, the reduction reveals localized hole solutions that are distinct from the classical Langer-Ambegaokar hole. In the Eckhaus-unstable regime, the same equation develops finite-time singularities, leading to phase slips. We derive a universal self-similar description of the approach to a phase slip, with parameter-independent scaling exponents in good agreement with direct numerical simulations. The associated similarity profile satisfies a nonlinear fourth-order boundary-value problem, for which we numerically identify multiple distinct solutions. Extending the reduction to the complex Ginzburg-Landau equation (CGLE) with almost real coefficients introduces an additional Burgers-type nonlinearity. This term produces wave steepening and traveling shocks connecting plane waves with different wave numbers. We derive the shock profile exactly and demonstrate numerically that solutions of the full CGLE converge to this prediction in the appropriate large-scale, near real-coefficient limit. Away from this asymptotic regime, the local wave number profile becomes nonmonotonic, an effect we explain using spatial dynamics. The shocks we discuss here are qualitatively distinct from the classical Nozaki-Bekki solutions. Our results thus identify a scalar reduced wave number equation that organizes several fundamental nonlinear phenomena in the Ginzburg–Landau equations, including stationary holes, self-similar collapse toward phase slips, and propagating shocks within a common large-scale description of pattern-forming systems.
\end{abstract}

\maketitle

\section{\label{sec:level1}Introduction}
The Ginzburg--Landau equation (GLE) is one of the most extensively studied nonlinear partial differential equations in physics, describing pattern formation in nonequilibrium systems near the onset of a pattern-forming instability \cite{newell1969finite,segel1969distant,Haken1983,cross1993pattern,aranson2002world,cross2009pattern,garcia2012complex}. Such patterns are characterized by an amplitude $R$ and a phase $\theta$. In the simplest case, a periodic pattern in one spatial dimension is specified by $R=R_0$ and $\theta=kx$, where $k$ is the constant pattern wave number. The GLE describes the slow evolution of the complex order parameter $u=R\exp i\theta$ on large spatial scales, i.e., the slow, large-scale modulation of both the pattern amplitude and its phase. When $R>0$ the number of cells in the domain is fixed although their width and location may change. Thus the cell number can only change when $R$ reaches zero at a point in the domain. At such a point the phase is undefined and the phase slips or jumps by $\pm 2\pi$, representing the injection or annihilation of a cell. In two dimensions such locations are associated with topological defects and the pattern undergoes a qualitative change when pairs of defects annihilate or are created. Thus phase slips are key to pattern relaxation, pattern coarsening \cite{CHE,gelens} and phase slip-driven front propagation \cite{Yiping}.

The GLE provides a simple model for studying the formation and evolution of a pattern due to phase slips and/or defect motion. Since the order parameter field is complex-valued, such studies usually require numerical simulations, even in one spatial dimension. It is therefore advantageous to reduce the problem to a single scalar equation for the local wave number of the pattern, $k\equiv\theta_x$. In this work we employ the WKB Ansatz to show that such a reduction of the GLE is possible and use the resulting nonlinear diffusion equation to study the formation of phase slips. The main contributions of the work are the following: 

\begin{itemize}
\item The nonlinear diffusion equation describes the evolution of the local wave number $k$ in the Eckhaus-unstable regime, i.e., when the pattern is modulationally unstable. We distinguish two cases, one arising from the GLE with real coefficients (RGLE) describing the Eckhaus instability for steady patterns \cite{eckhaus1965studies,kramer1985eckhaus,hoyle2006pattern}, and the other from the GLE with complex coefficients (CGLE) describing traveling waves \cite{janiaud}. 
\item We use the nonlinear diffusion equation to construct a family of exact stationary localized states in the RGLE case, and traveling fronts in the CGLE case, and identify their counterparts in the underlying GLE. 
\item We demonstrate explicitly that the approach to a phase slip in the nonlinear diffusion equation is self-similar, and solve the resulting nonlinear eigenvalue problem numerically, thereby substantially extending existing studies of phase slips \cite{kramer1985eckhaus,eckmann1995phase}.
\item In the CGLE case our equation reduces to the Kuramoto-Sivashinsky equation \cite{kuramoto1976persistent,kuramoto1976turbulent,sivashinski1977nonlinear,michelson1977nonlinear,kuramoto1978diffusion,kuramoto1984chemical,sivashinsky1985weak} when the modulation wave number is small. Thus our equation represents a generalization of the Kuramoto-Sivashinsky equation.
\item We find that in the Eckhaus-unstable regime phase slips occur whenever the system does not relax to a stable, spatially homogeneous state, suggesting that phase turbulence \cite{aranson2002world} is a transient state.
\item We validate our results against direct numerical simulations (DNS) of the wave number equation and the GLE.
    \end{itemize}

This paper is organized as follows. In Sec.~\ref{sec:derivation} we derive the nonlinear diffusion equation for the local wave number $k(x,t)$ from the complex GLE. In Sec.~\ref{sec:RGLE} we analyze this equation in the  RGLE case, identifying its gradient structure and constructing exact stationary solutions, which are shown to be unstable and distinct from the Langer-Ambegaokar stationary hole solution \cite{langer1967intrinsic,lega2001traveling}. In Sec.~\ref{sec:self_similar}, we describe the universal self-similar dynamics governing the approach to finite-time singularities in the wave number equation and validate these predictions numerically. Section~\ref{sec:cgle} concerns the complex GLE. The paper concludes with a discussion in Sec.~\ref{sec:conclusions}.

\section{Derivation of the reduced wave number equation \label{sec:derivation}}
We consider the Ginzburg-Landau equation (GLE) with (weakly) complex coefficients 
\begin{equation}
\epsilon^2u_\tau
=
u+(1+i\epsilon \alpha )\epsilon^2 u_{\xi\xi}
-(1+i\epsilon\beta)|u|^2u,
\label{eq:cgl}
\end{equation}
written in slow spatial and temporal variables $\xi=\epsilon x$ and $\tau=\epsilon^2 t$, where $0<\epsilon\ll 1$.
The factor $\epsilon^2$ multiplying the highest spatial derivative motivates a rapidly oscillating (WKB) approximation
\cite{knobloch2014stability},
\begin{equation}
u(\xi,\tau)=R(\xi,\tau)\exp\!\left(\frac{i}{\epsilon}\phi(\xi,\tau)\right),
\label{eq:wkb_ansatz}
\end{equation}
in terms of a real amplitude $R$ and phase $\phi$, with local wave number $k\equiv\phi_\xi=O(1)$ and $R=O(1)$. In contrast to classical weakly nonlinear theory, we do not assume $|u|\ll1$. Instead, we consider fully nonlinear wavetrains whose modulation occurs on long spatial and temporal scales. At leading order, the amplitude is slaved to the wave number,
\begin{equation}
R=\sqrt{1-k^2},
\label{eq:amp}
\end{equation}
so that $|k|\le1$. Phase slips correspond to $R\to0$, which occurs as $k\to\pm1$, a limit in which the WKB expansion loses uniform validity. 

At $O(\epsilon)$ one obtains 
\begin{equation}
k_\tau
=
\partial_\xi\!\left[D(k)k_\xi\right]
+
2(\beta-\alpha)kk_\xi,
\label{eq:nl_diffusion_equation_complex_2nd_order}
\end{equation}
with the state-dependent diffusion coefficient
\begin{equation}
D(k)=\frac{1-3k^2}{1-k^2},
\end{equation}
which becomes negative  at $|k|>1/\sqrt{3}$, signaling the onset of the Eckhaus instability
\cite{eckhaus1965studies,kramer1985eckhaus,raitt1995domain,hoyle2006pattern}.
For $\alpha=\beta$, Eq.~\eqref{eq:nl_diffusion_equation_complex_2nd_order} reduces to a nonlinear diffusion equation closely related to generalized porous-medium models
\cite{pattle1959diffusion,heaslet1961diffusion,shampine1973concentration,aronson2006porous}. The validity of the corresponding long-wavelength phase-diffusion approximation for the RGLE has been rigorously established for sufficiently smooth initial conditions in the Eckhaus-stable regime \cite{melbourne2004phase}. Moreover, the same diffusion coefficient arises from linear stability analysis of periodic wavetrains with wave number $k_0$, yielding
$\sigma(q)=-q^2D(k_0)+O(q^4)$ \cite{kramer1985eckhaus}, where $D(k_0)<0$ indicates the Eckhaus instability.
When $\alpha\neq\beta$, a Burgers-type quadratic nonlinearity is present, leading to gradient steepening and shock formation.

Equation~\eqref{eq:nl_diffusion_equation_complex_2nd_order} provides a leading-order description of the slow phase dynamics. Higher-order asymptotics generate regularizing contributions involving higher-order derivatives. For practical purposes, we model these by adding a constant-coefficient hyperdiffusive term ($\kappa>0$),
\begin{align}
k_\tau
&=
\partial_\xi\!\left[D(k)k_\xi\right]
+2(\beta-\alpha)kk_\xi
-\kappa k_{\xi\xi\xi\xi},
\label{eq:nl_diff_eq_full}
\end{align}
which we use throughout the remainder of this work. With this term, Eq.~\eqref{eq:nl_diff_eq_full} resembles the convective Cahn-Hilliard equation studied in \cite{emmott,golovin,watson,podolny};
in particular, the two equations share the reflection symmetry ${\cal R}: (\xi,k)\to -(\xi,k)$. 

A systematic $O(\epsilon^2)$ WKB reduction described in Appendix~\ref{app:hi_ord_wkb} generates multiple higher-order regularizing contributions with up to fourth-order derivatives of the wave number. Specifically for the RGLE ($\alpha=\beta=0)$, neglecting terms smaller than $O(\epsilon^2)$ one finds
\begin{equation}
k_\tau=
\partial_\xi\!\big(D(k)k_\xi\big)
-\epsilon^2\partial_\xi
\Big(
E(k)k_{\xi\xi\xi}
+F_1(k)k_\xi k_{\xi\xi}
+F_2(k)k_\xi^3
\Big),\label{eq:higher_order_wkb_main_text} 
\end{equation}
where $E(k)$, $F_1(k)$ and $F_2(k)$ are explicit functions of the local wave number which become singular as $k\to 1$. Thus the  constant-coefficient hyperdiffusive regularization should be viewed as a Kuramoto-Sivashinsky-like minimal model capturing short-scale regularization rather than as an exact asymptotic formulation. In the interest of simplicity, we focus in the remainder of this paper on Eq.~\eqref{eq:nl_diff_eq_full}, leaving the analysis of Eq.~\eqref{eq:higher_order_wkb_main_text} for future study.

\section{Real Ginzburg-Landau equation (Or $\alpha=\beta$)  } \label{sec:RGLE}
We begin by considering the real-coefficient Ginzburg-Landau equation, for which $\alpha=\beta=0$, noting that Eq.~(\ref{eq:nl_diff_eq_full}) predicts the same behavior for the CGLE whenever $\alpha=\beta\ne0$. The real-coefficient case is important because its simpler mathematical structure is particularly amenable to theoretical analysis, as described below.

\subsection{Variational Structure}
Equation~(\ref{eq:cgl}) with $\alpha=\beta=0$ possesses variational structure:
\begin{equation}
    \partial_\tau u = - \frac{\delta \mathcal{F}_{RGLE}}{\delta \overline{u}},
    \label{variational}
\end{equation}
where $\delta/\delta \overline{u}$ denotes the variational derivative with respect to the complex conjugate of $u$, with the free energy functional
\begin{equation}
    \mathcal{F}_{RGLE}[u]= \frac{1}{\epsilon^2}\int \left(\frac{\epsilon^2}{2} |u_\xi|^2 -\frac{1}{2}|u|^2 + \frac{1}{4}|u|^4 \right)d\xi.
\end{equation}
This implies that 
\begin{equation}
    \partial_\tau \mathcal{F}_{RGLE} = - \int u_\tau^2 d\xi \leq 0,
\end{equation}
where equality holds for stationary states only. Given this variational structure of the RGLE, it is natural to inquire whether Eq.~(\ref{eq:nl_diff_eq_full}) with $\alpha=\beta$ can similarly be expressed in terms of a suitable free energy. The reduction of the GLE to Eq.~(\ref{eq:nl_diff_eq_full}) is subtle because it eliminates the amplitude degree of freedom by slaving it to the wave number, transforming the complex order parameter GLE into an evolution equation for a real-valued wave number whose gradient-flow structure is therefore inherited in a nontrivial way. 

Using the amplitude-phase variables $R,\phi$ from Eq.~\eqref{eq:wkb_ansatz}, without making any
approximation, one finds after some algebra that $R^2 \phi_\tau = \partial_\xi(R^2\phi_\xi)
= - \delta \mathcal{F}_{RGLE}/\delta \phi$. Since $k=\phi_\xi$, the wave number obeys a conservation law. Moreover, the reduced equation, Eq.~(\ref{eq:nl_diffusion_equation_complex_2nd_order}), possesses a Cahn-Hilliard-like conserved gradient structure
\begin{equation}
    k_\tau = \partial_{\xi \xi} \frac{\delta\mathcal{F}}{\delta k}
    \equiv \partial_{\xi\xi} \mu,
    \label{eq:conserved_gradient_structure}
\end{equation}
where the free energy (Lyapunov) functional $\mathcal{F}$ is
distinct from $\mathcal{F}_{RGLE}$ and, upon inclusion of the
hyperdiffusive regularization, is given by
\begin{equation}
    \mathcal{F}[k] = \int \left(f(k)
    +\frac{\kappa}{2}|k_\xi|^2\right) d\xi.
    \label{eq:def_free_energy}
\end{equation}
Here $\delta\mathcal{F}/\delta k\equiv \mu=f'(k)-\kappa k_{\xi\xi}$
is the chemical potential and the local free energy density $f(k)$ satisfies $f''(k)=D(k)=(1-3k^2)/(1-k^2)$. We stress the importance of the conserved structure of the variational principle for the wave number in Eq.~\eqref{eq:conserved_gradient_structure}, in contrast to Eq.~\eqref{variational} for $u$. Direct integration yields
\begin{align}
    f'(k) =& \,3k +\ln\left(\frac{1-k}{1+k}\right) +\frac{C_1}{2},\\
    f(k)  = &\, k\ln\left(\frac{1-k}{1+k}\right)-\ln(1-k^2)\nonumber\\
     &+\frac{3}{2}k^2+\frac{C_1 }{2}k+\frac{C_2}{2},
\end{align}
where $C_1, C_2$ are constants of integration and we have used
the fact that $|k|<1$.

With suitable boundary conditions (e.g., zero-flux boundaries, or periodic or infinite domains with suitable decay at infinity), Eq.~(\ref{eq:conserved_gradient_structure}) implies that the free energy decreases monotonically
\begin{equation}
    \frac{d \mathcal{F}}{d \tau} = \int \mu k_\tau\, d\xi
= \int \mu\, \partial_{\xi\xi}\mu\, d\xi = - \int (\mu_\xi) ^2 d\xi\leq 0,
\end{equation}
with equality when $\mu_\xi\equiv 0$, i.e., in the equilibrium state. This indicates that the dynamics are irreversible, and no sustained oscillatory dynamics are possible, with the long-time behavior being constrained by the energy landscape.

Integrating Eq.~(\ref{eq:nl_diff_eq_full}) in $\xi$ over the entire domain gives
\begin{equation}
    \frac{\rm d}{\rm d \tau}\langle k \rangle \equiv \frac{\rm d}{\rm d \tau} \int k d\xi 
 = -J|_-^+= \left[ D(k) k_\xi - \kappa k_{\xi\xi\xi}\right]_-^+,
\end{equation}
where the right-hand side vanishes for zero flux boundary conditions $k_\xi=k_{\xi\xi\xi} = 0$, which we refer to as \textit{free-free boundary conditions}, as well as for periodic boundary conditions or asymptotically uniform states on an infinite domain. On a finite domain, the resulting conservation law $\langle k \rangle =const. = \Delta \phi$ (where $\Delta \phi$ is the change in phase across the domain) rules out traveling fronts of fixed shape connecting distinct values of $k$, regardless of domain size.

\subsection{Linear dispersion relation}
\begin{figure}
    \centering
    \includegraphics[width=\linewidth]{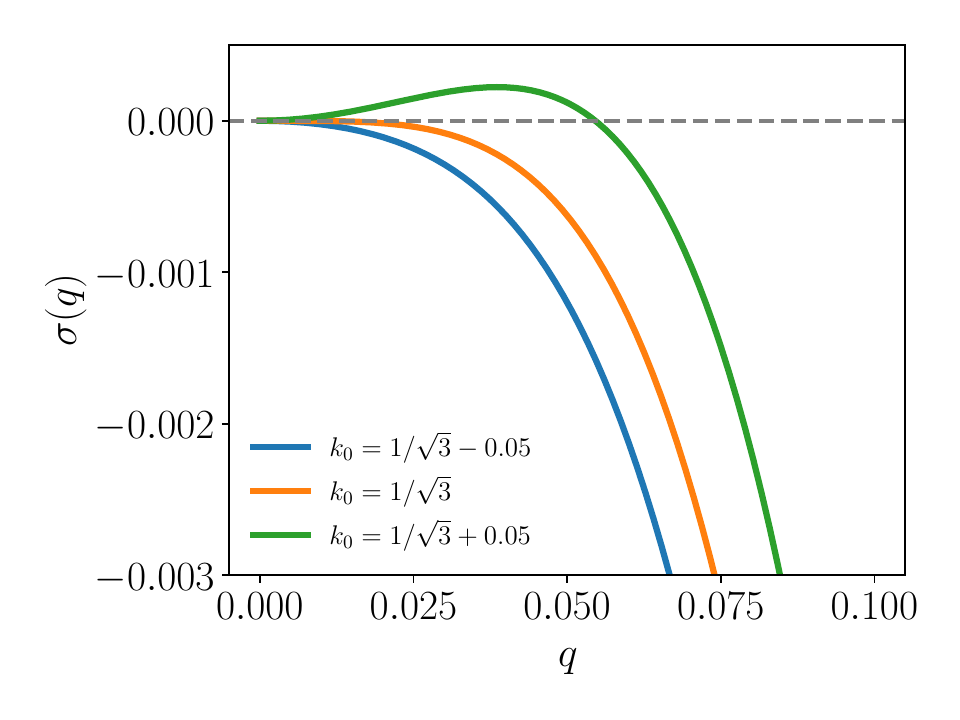}
    \caption{Dispersion relation for different background wave numbers $k_0$ in terms of the growth rate $\sigma(q)$, where $q$ is the perturbation wave number. Inside the Eckhaus boundary $k_0=1/\sqrt{3}$, there is no instability: the growth rate $\sigma(q)<0$ for all $|q|>0$ and vanishes at $q=0$. When the Eckhaus boundary is crossed, $k_0>1/\sqrt{3}$, the regularizing fourth-order term suppresses instability at large $q$ but cannot prevent instability at small $q$. }
    \label{fig:disp_rel_ill}
\end{figure}

The fourth-order term has a stabilizing effect with respect to small-scale perturbations. For uniform states $k=k_0$, this can easily be seen in the growth rate $\sigma(q)$ of perturbations with wave number $q$,
\begin{equation}
    \sigma(q) = -D(k_0)q^2  -\kappa q^4, \label{eq:disp_rel}
\end{equation}
where $D(k_0)=(1-3k_0^2)/(1-k_0^2)>0 $. This growth rate is negative for all $q$ when $|k_0|<1/\sqrt{3}$ (inside the Eckhaus boundary), implying that $k=k_0$ is linearly stable, while for $|k_0|>1/\sqrt{3}$ the growth rate is positive at small $q$ provided the domain is large enough to accommodate these large scales, as illustrated in Fig.~\ref{fig:disp_rel_ill}.
\subsection{Critical points of the free energy: localized pulse and periodic modulations}
\label{sec:steady_solutions}
\begin{figure}
    \centering
    \includegraphics[width=\linewidth]{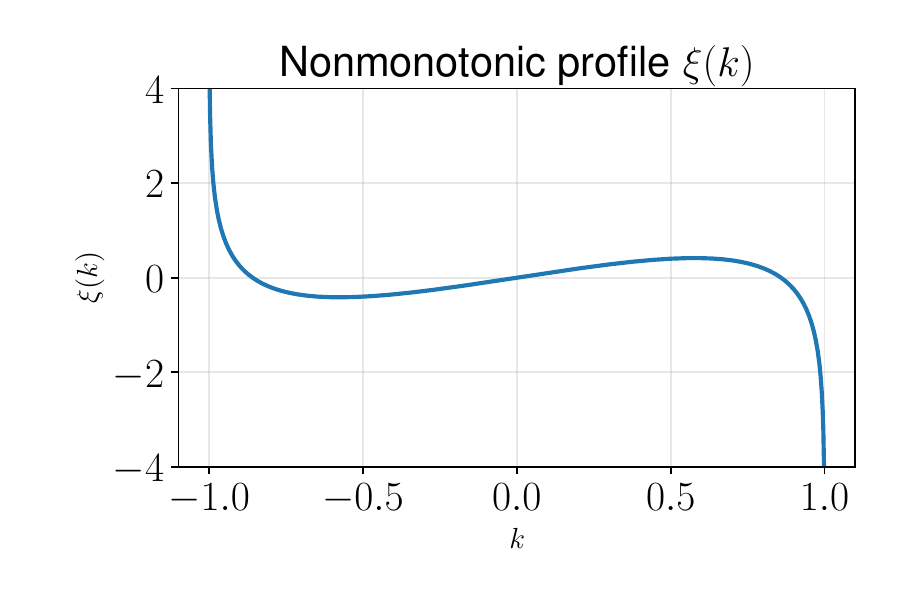}
    \caption{ In the absence of the stabilizing hyperdiffusive term, one finds a noninvertible profile $\xi(k)$ following Eq.~(\ref{eq:failed_stat_sol_diffusion_only}) for $A=1$, $B=0$, illustrating that no smooth stationary solutions $k(\xi)$ can be found if only the nonlinear diffusion term is considered. }
    \label{fig:xi_of_k_noninvertible}
\end{figure}

We begin by seeking time-independent solutions $k\equiv k(\xi)$, which correspond to critical points of $\mathcal{F}[k]$,
\begin{eqnarray}D(k)k_\xi-\kappa k_{\xi\xi\xi}=A \nonumber \\ \Leftrightarrow \, k_\xi\left(D(k)-\kappa\frac{dk_{\xi\xi}}{dk}\right)=A,\label{eq:1st_integral}
\end{eqnarray}
where again $D(k)=(1-3k^{2})/(1-k^{2})$ 
and we view $k_{\xi\xi}$ as a function of $k$. For $\kappa=0$, Eq.~\eqref{eq:1st_integral} integrates to 
\begin{eqnarray}
3k+\ln[(1-k)/(1+k)]=A\xi+B. \label{eq:failed_stat_sol_diffusion_only}
\end{eqnarray}
 If $A\neq0$, no smooth solution can cross the zeros of $D(k)$ at $k=\pm1/\sqrt3$, where $k_\xi=A/D(k)$ diverges, and the implicit relation cannot define a smooth profile on the entire real line, as illustrated in Fig.~\ref{fig:xi_of_k_noninvertible}. If $A=0$, all smooth stationary solutions are spatially uniform. Thus, nonlinear diffusion alone cannot support globally smooth, nonuniform stationary profiles.

In the following, unless stated otherwise, we take $\kappa>0$. This allows for the existence of nonuniform stationary states, since the stabilizing effect of hyperdiffusion can balance the destabilizing influence of antidiffusion. Such solutions are known to exist in the real Ginzburg-Landau equation, for instance, the Langer-Ambegaokar hole solution \cite{langer1967intrinsic,lega2001traveling}. On an infinite domain, we can rescale space and time to eliminate $\kappa$ from the equation and in the following we therefore set $\kappa= 1$.

We consider solutions $k(\xi)$ of Eq.~(\ref{eq:1st_integral}) which become $\xi$-independent at $\xi\to\pm \infty$, which requires $A=0$. One possible solution of (\ref{eq:1st_integral}) is again given by $k_\xi=0$, i.e., a uniform state. Nonuniform, i.e., $\xi$-dependent solutions obey
\begin{equation}
    D(k)=f''(k)=\frac{d k_{\xi\xi}}{d k} \,,\label{eq:nonuniform_critical_points1}
\end{equation}
where $f(k)$ is the local free energy density introduced in Eq.~(\ref{eq:def_free_energy}). Integrating in $k$ gives
\begin{equation}
     k_{\xi\xi}=f'(k)=3k +\ln\left(\frac{1-k}{1+k}\right) +\frac{C_1}{2} \,\label{eq:nonuniform_critical_points2}
\end{equation}
(assuming $|k|<1$ again), which is of the form of Newton's second law with unit mass, $\xi$ corresponding to `time' and $-f(k)$ corresponding to the potential. Multiplying by $k_\xi$ and integrating over $\xi$ gives
\begin{align}
     k_\xi^2=2f(k)&= 2\left[k\ln\left(\frac{1-k}{1+k}\right)-\ln(1-k^2)\right]\nonumber\\
     &\,\,\,\,\,\,\,\,+3k^2+C_1k+C_2 \,\label{eq:nonuniform_critical_points3}\\
     &\equiv g(k) + C_1k + C_2. \nonumber
\end{align}
This result is the `energy conservation' relation associated with Eq.~(\ref{eq:nonuniform_critical_points2}). 

Depending on the choice of $C_1$, $C_2$, one of two types of nonconstant solutions may arise on an unbounded domain: localized homoclinic solutions and spatially periodic solutions. Their properties are summarized below.

\begin{figure}[hbt!]
    \centering    \includegraphics[width=1\linewidth]{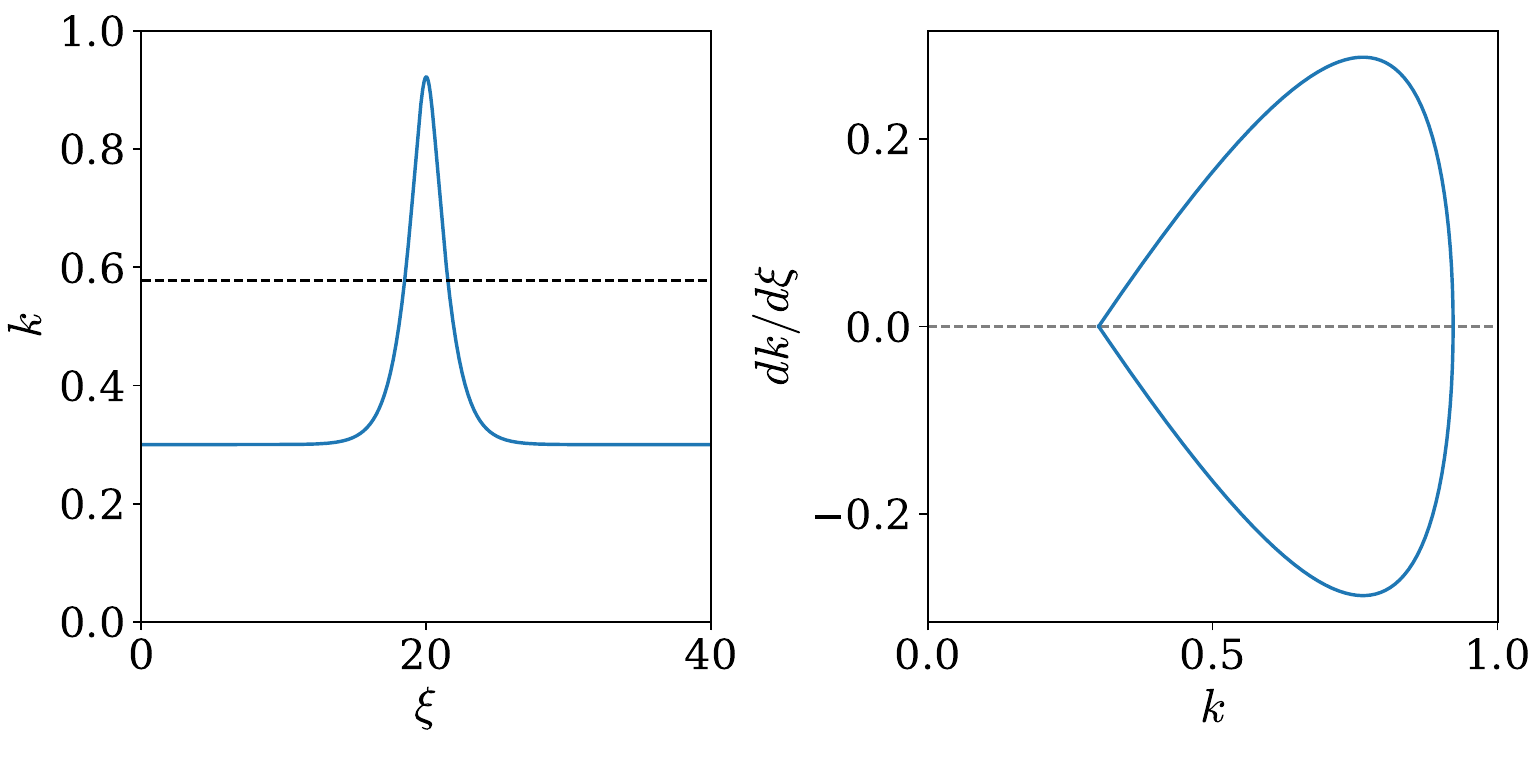}
    \caption{Example of a pulse solution starting and ending at $k=0.3$ at $\xi\to \pm \infty$. Left: spatial wave number profile straddling the Eckhaus boundary, indicated by the horizontal dashed line. Right: corresponding phase space portrait.}
    \label{fig:pulse}
\end{figure}

\begin{figure}
    \centering
    \includegraphics[width=\linewidth]{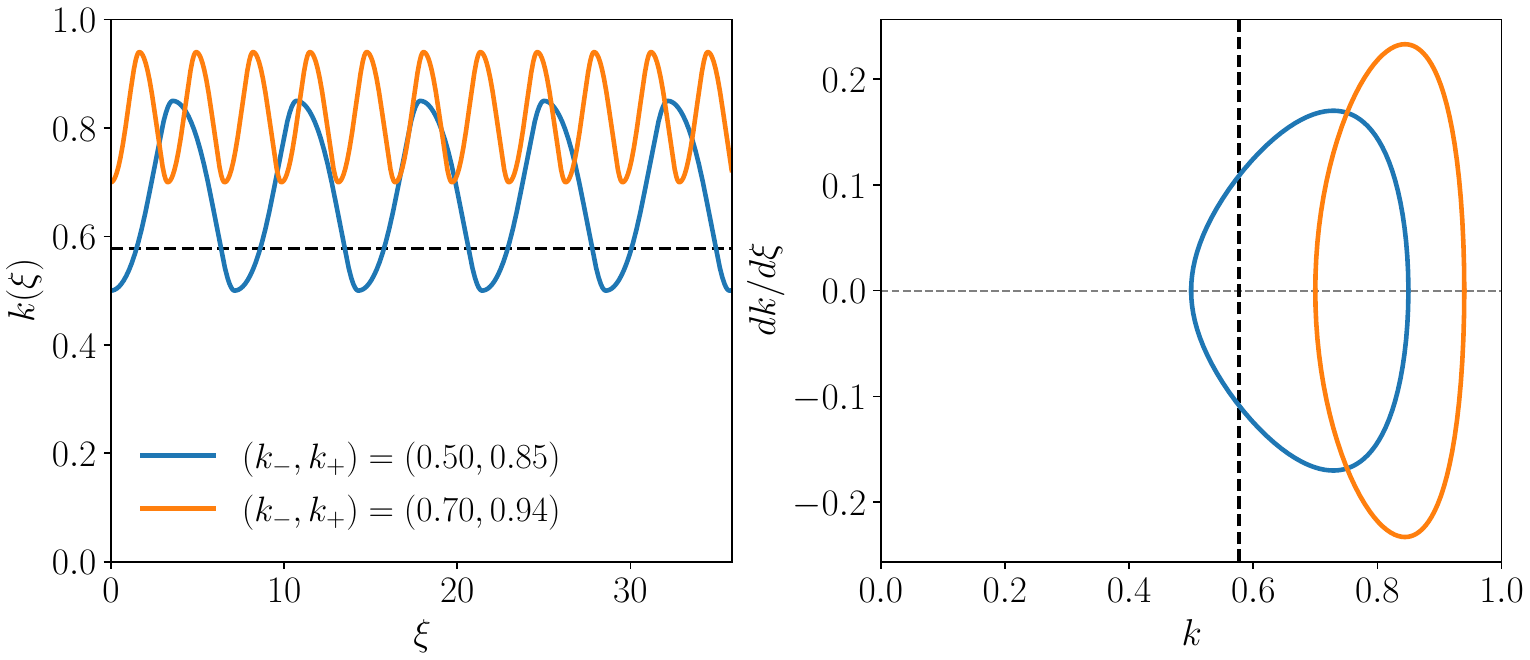}
    \caption{Stationary, spatially extended, periodic modulations in the pattern wave number (and amplitude) between $k=k_-$ and $k=k_+$. Left panel: spatial profile. Right panel: phase space portrait. Positivity of $k_\xi^2=2f(k)$ prevents such states from existing entirely within the Eckhaus-stable region: the maximum wave number in a period must lie within the Eckhaus-unstable range, while the minimum may lie either within the Eckhaus-stable or within the unstable interval.}
    \label{fig:extended_patterns}
\end{figure}

\begin{enumerate}
\item Equations~(\ref{eq:nonuniform_critical_points1})--(\ref{eq:nonuniform_critical_points3}) admit no smooth heteroclinic orbit connecting two distinct asymptotic wave numbers $k_1\neq k_2$ as $\xi\to\pm\infty$, and hence neither does Eq.~\eqref{eq:nl_diff_eq_full} for $\alpha=\beta=0$. \textbf{Proof}: see Appendix~\ref{app:proofs}. 
\item The only nonconstant homoclinic orbit smoothly connecting a given wave number $k_\infty\in[0,1/\sqrt{3})$ at $\xi\to-\infty$ to the same wave number at $\xi\to\infty$ has a single local maximum whose value exceeds $1/\sqrt{3}$, i.e., the solution straddles the Eckhaus stability boundary (cf. Fig.~\ref{fig:pulse}). By the symmetry $k\to-k$, the corresponding solutions for negative $k_\infty$ have a single local minimum below $-1/\sqrt{3}$. \textbf{Proof}: see Appendix~\ref{app:proofs}.

\item The nonconstant homoclinic orbit described above is necessarily linearly unstable. \textbf{Proof}: see Appendix~\ref{app:proofs}.

\item In addition to the localized homoclinic orbit discussed above,
Eqs.~(\ref{eq:nonuniform_critical_points1})--(\ref{eq:nonuniform_critical_points3}) also admit spatially periodic stationary solutions, corresponding to extended modulations of the wave number. Such solutions arise when $f(k)$ has two consecutive simple zeros $k_-<k_+$, with $k_\xi^2=2f(k)>0$ for $k\in(k_-,k_+)$ and $f'(k_-)>0$, $f'(k_+)<0$.
The resulting profile $k(\xi)$ oscillates periodically between
these turning points. Since $f''(k)>0$ in the Eckhaus-stable range, the positivity constraint on $f(k)$ implies that the turning points must either both lie in the Eckhaus-unstable region or straddle the boundary between the Eckhaus-stable and Eckhaus-unstable regions, cf. Fig.~\ref{fig:extended_patterns}.

In contrast, localized homoclinic solutions require a double zero of $f(k)$ at the asymptotic wave number $k_\infty$, satisfying $f(k_\infty)=f'(k_\infty)=0$, which permits exponential convergence $k\to k_\infty$  as $|\xi|\to\infty$.

\item The spatially periodic steady states described above are linearly unstable independently of the choice of $k_+, k_-$.
\textbf{Proof}: see Appendix~\ref{app:proofs}.

\end{enumerate}

To summarize the above results, in addition to spatially uniform states, the only type of smooth homoclinic, localized stationary solutions consists of a single pulse in the wave number, asymptoting to a wave number in the Eckhaus-stable regime at $\xi\to \pm \infty$, while necessarily peaking within the Eckhaus-unstable regime. Such spatially localized homoclinic pulses (also known as ``reversal waves'' or ``spikes'') are shown to be linearly unstable in Eq.~\eqref{eq:nl_diff_eq_full} with $\alpha=\beta$. In addition, there exist stationary periodic solutions of two types: they either straddle the Eckhaus boundary or lie entirely in the Eckhaus unstable regime. These periodic solutions are also linearly unstable. The linear instability of these nontrivial stationary states is not surprising, given that their profiles extend into the Eckhaus-unstable regime. This observation provides a physical intuition for their instability, although it does not in itself constitute a proof. Furthermore, similar results have been established rigorously for the standard Cahn-Hilliard equation \cite{Howard2009SpectralCHstationary,Howard2011SpectralPeriodicCH}, and nonmonotonic solutions such as localized pulses are known to be unstable in scalar reaction-diffusion systems \cite{Hagan1981InstabilityNonmonotonic}, including in the presence of a conservation law \cite{PoganScheel2010InstabilitySpikes,PoganScheel2013RadialSpikes}. In fact, the instability of localized and periodic stationary solutions was established in a broader class of quasi-gradient systems \cite{PoganScheelZumbrun2013QuasiGradient}. In most of the works cited above, the spectral stability properties of stationary solutions were established using the Evans function, see also \cite{oh2003stability,howard2004evans,kapitula2005stability,barker2018evans}. In contrast, the demonstrations we present in Appendix \ref{app:proofs} rely on classical results about Schrödinger operators naturally appearing in the linearized problem here. The Evans function is not utilized in our analysis.

We have performed direct numerical simulations using Dedalus \cite{burns2020dedalus} confirming the instability of these solutions on a finite computational domain subject to the free-free boundary conditions $k_\xi=k_{\xi\xi\xi}=0$ at either end or with periodic boundary conditions (not shown, see also Sec.~\ref{sec:validation_ss_dns}). The only stable states on $\mathbb{R}$ are those of uniform wave number in the Eckhaus-stable regime. 

In order to explicitly construct the solutions shown in Figs.~\ref{fig:pulse} and \ref{fig:extended_patterns}, one needs to determine suitable values for the constants of integration $C_1,C_2$. One approach is to impose the values of $k$, denoted by $k_a<k_b$, at the ends of the interval of interest, which terminates at $k_\xi(k=k_a)=k_\xi(k=k_b)=0$. Equation~\eqref{eq:nonuniform_critical_points3} gives two conditions which can be solved to find $C_1=[g(k_b)-g(k_a)]/(k_a-k_b)$ and $C_2=-g(k_a)-C_1 k_a$. Generically, this results in an interval of finite extent in $\xi$, as in Fig.~\ref{fig:extended_patterns}. To obtain the homoclinic pulse solution in Fig.~\ref{fig:pulse}, which extends over $\xi \in \mathbb{R}$, one may prescribe the wave number at infinity using $k_\xi(k=k_a)=0$ and additionally impose $k_{\xi\xi}(k=k_a)=0$ utilizing Eqs.~\eqref{eq:nonuniform_critical_points2} and \eqref{eq:nonuniform_critical_points3}, which yields $C_1 = 2\left\lbrace3k_a - \ln[(1-k_a)/(1+k_a)]\right\rbrace$ and $C_2=-F(k_a) - C_1 k_a$. The maximum wave number $k_b$ attained at the peak can be found by solving $k_\xi(k=k_b)=0$. Symmetry is then imposed to construct the full profile. 

Figure~\ref{fig:pulse} illustrates the solution consisting of a single pulse in terms of $k$ versus $\xi$ (left panel) and in the phase space spanned by $k$ and $dk/d\xi$ (right panel), 
revealing a homoclinic orbit in the spatial dynamics picture. The solution is of hole type, i.e., it corresponds to a local depression in the pattern amplitude, while the wave number is locally increased -- similar states have been observed, e.g., in thermosolutal convection \cite{spina1998confined}.

\subsection{Comparison with Langer--Ambegaokar hole solution}
For the RGLE, i.e., $\alpha=\beta=0$ in Eq.~\eqref{eq:cgl}, Langer and Ambegaokar constructed
a one-parameter family of stationary phase-winding (hole) solutions
\cite{langer1967intrinsic,lega2001traveling}. In the notation of Eq.~\eqref{eq:cgl},
a stationary hole takes the form
\begin{equation}
u(\xi)=\sqrt{1-q^2}\,
\exp\!\Bigl(i\,\frac{q}{\epsilon}\,\xi\Bigr)\,B(\xi),
\label{eq:LA_hole_scaled}
\end{equation}
where 
\begin{equation}
B(\xi)=\frac{q+i\kappa\,\tanh\!\bigl(\kappa \xi/\epsilon\bigr)}{q+i\kappa}.
\label{eq:LA_B}
\end{equation}
and $q$ and $\kappa$ are real parameters satisfying
\begin{equation}
1=3q^2+2\kappa^2.
\label{eq:LA_constraint}
\end{equation}

This solution, stated incorrectly in \cite{lega2001traveling}, represents a localized amplitude depression on a constant wave number
background wave, with $|B(\xi)|\to1$ as $\xi\to\pm\infty$. Such a homoclinic hole structure is sometimes referred to as a homoclon \cite{aranson2002world}.

\begin{figure} \centering \includegraphics[width=\linewidth]{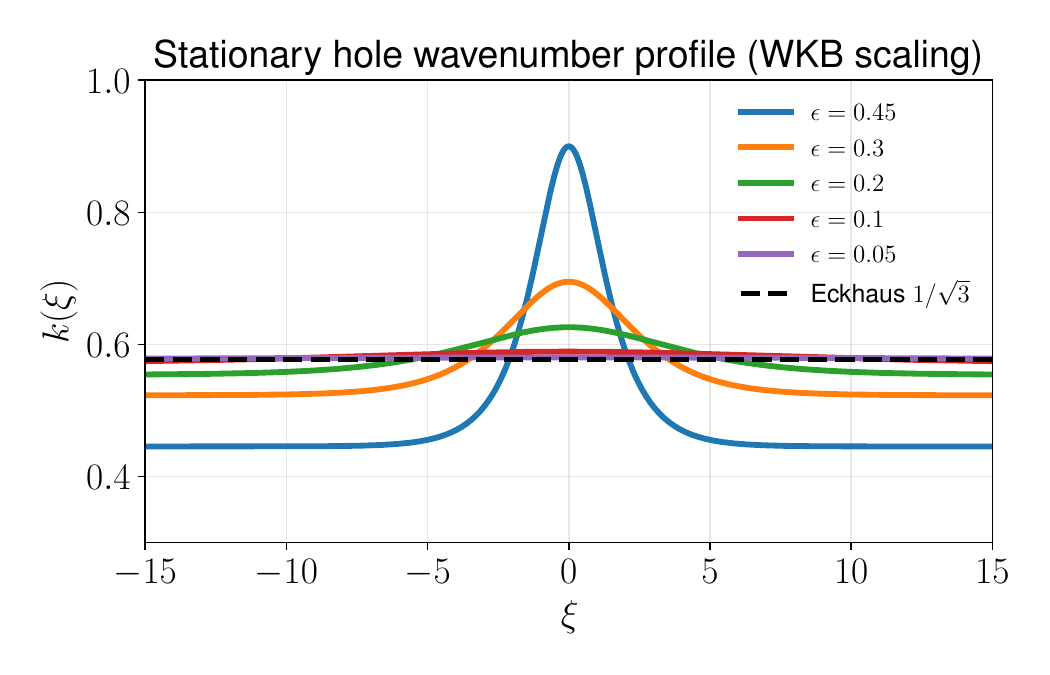} \caption{Local wave number $k(\xi)$ associated with the stationary RGLE hole in Eq.~\eqref{eq:LA_hole_scaled} for decreasing $\epsilon$. As $\epsilon\to0$, the profile collapses to the uniform Eckhaus-critical value $k=1/\sqrt{3}$, illustrating the degeneration of the hole into a plane wave in this limit.} \label{fig:illustr_standing_hole} \end{figure}

Two immediate consequences follow from Eqs.~\eqref{eq:LA_B} and~\eqref{eq:LA_constraint}. First, an $O(1)$ width defect in $\xi$ requires $\kappa=\epsilon\lambda$ with $\lambda=O(1)$. Second, under this scaling, Eq.~\eqref{eq:LA_constraint} yields $q=1/\sqrt{3}+O(\epsilon^2)$, i.e. any stationary hole that remains $O(1)$ wide in the slow variable necessarily lies asymptotically close to the Eckhaus boundary. Accordingly, the associated local wave number
$k(\xi)=\epsilon \partial_\xi \arg u(\xi)$ has vanishing $O(1)$ modulation as $\epsilon\to0$, and the solution collapses to the uniform Eckhaus-critical plane wave, as illustrated in
Fig.~\ref{fig:illustr_standing_hole}.

This behavior contrasts sharply with the localized homoclinic pulse solutions of the nonlinear diffusion equation discussed above, which generically exhibit $O(1)$ variations of $k(\xi)$ over $O(1)$ spatial intervals and necessarily penetrate the Eckhaus-unstable regime. We therefore conclude that the stationary RGLE hole \eqref{eq:LA_hole_scaled} is qualitatively distinct from our homoclinic pulse: in the reduced description the former collapses to an Eckhaus-critical plane wave, whereas our pulse remains genuinely localized in the wave number field. Only in the limit $k_\infty\to  1/\sqrt{3}$ do the two profiles coincide. Another interesting observation is that the hole solution in Eq.~\eqref{eq:LA_hole_scaled} is unstable, a finding confirmed using pseudospectral direct numerical simulations (DNS) in a periodic domain (not shown) -- initializing with this solution leads to decay to a uniform wave number in our simulations.

\section{Self-similar evolution towards finite-time singularities in the reduced equation}
\label{sec:self_similar}
The dynamics of phase slips has been studied in detail in \cite{kramer1985eckhaus, aranson2002world,eckmann1995phase}. Here we study the corresponding dynamics within the reduced description (\ref{eq:nl_diff_eq_full}).

\subsection{Theoretical analysis}

For certain solutions of Eq.~(\ref{eq:nl_diff_eq_full}) entering the Eckhaus-unstable regime, the local wave number approaches $k\to 1^-$ in finite time. This corresponds to a singularity in the WKB-derived diffusion coefficient $D(k)$ and physically represents the approach to a phase slip, where the pattern amplitude $R\to 0$ at an isolated point. Below, we analyze the dynamics of the reduced equation for $k$ close to $1$. We define $q\equiv 1-k$, which we assume to be small and positive. Taylor-expanding in $q\ll 1$ gives $D(k)\approx-q^{-1}+O(1)$, and Eq.~(\ref{eq:nl_diff_eq_full}) becomes, at leading order,
\begin{equation}
q_\tau=
-\frac{\partial}{\partial\xi}\left(q^{-1}q_\xi\right)
-\kappa q_{\xi\xi\xi\xi}
+2(\alpha-\beta)q_\xi .
\label{q-eq}
\end{equation}

To identify the dominant balances in the spatio-temporal dynamics near the phase slip, we seek a similarity solution of the form
\begin{equation}
q(\xi,\tau)=|\tau_*-\tau|^a H(\eta),
\qquad
\eta=\frac{\xi}{|\tau_*-\tau|^b},
\end{equation}
where $a,b>0$ are as yet unknown constants, and the phase slip is assumed to occur at time $\tau=\tau_*$ and location $\xi=0$. Solutions of this type have been studied extensively in the theory of nonlinear diffusion equations \cite{grundy1979similarity} and numerous other partial differential equations \cite{huppert1982propagation,rottman1983gravity,gratton1994self}, in particular in the context of singularity formation \cite{eggers2008role}.

Substituting the similarity ansatz into Eq.~(\ref{q-eq}), we find
\begin{widetext}
\begin{equation}
-aH+b\eta H'
=
-s^{1-2b-a}\left(\frac{H'}{H}\right)'
-\kappa s^{1-4b}H''''
+2(\alpha-\beta)s^{1-b}H',
\label{eq:ss_all_terms}
\end{equation}
\end{widetext}
where we denote $s\equiv |\tau_*-\tau|$ for brevity. In the limit $s\to 0$, the dominant balance between the left-hand side and the nonlinear diffusion and hyperdiffusion terms on the right-hand side requires $a=1-2b$ and $1-4b=0$. Hence
\begin{equation}
a=\frac12, \qquad b=\frac14
\end{equation}
and the leading-order similarity equation is
\begin{equation}
-\frac12H+\frac14\eta H'
=-(H'/H)'-\kappa H''''.
\label{eq:nlbvp}
\end{equation}
The hyperdiffusion coefficient $\kappa>0$ is set to unity by the rescaling $\eta=\kappa^{1/4}\tilde{\eta}$ and $H=\kappa^{-1/2}\tilde{H}$, after which tildes are dropped for simplicity. Equation~\eqref{eq:nlbvp} bears a strong resemblance to the fourth-order similarity equation derived by Zhang and Lister for thin-film rupture \cite{ZhangLister1999}, although the similarity exponents are different and the fourth-order regularizing term in their problem is nonlinear.

The symmetry of Eq.~\eqref{eq:nlbvp} under $\eta\to-\eta$ allows us to consider the problem on the half-line $0\leq\eta<\infty$. Useful theoretical insight can be obtained from the large-$\eta$ behavior of $H(\eta)$. Substitution of the expansion
\begin{equation}
H(\eta)\sim
C\eta^2+A+B\eta^{-2}+D\eta^{-4}
+O(\eta^{-6}) \label{eq:asymp_exp}
\end{equation}
into Eq.~\eqref{eq:nlbvp} gives
\begin{equation}
A=0,\qquad B=-2,\qquad D=0,
\end{equation}
and therefore
\begin{equation}
H(\eta)\sim C\eta^2-2\eta^{-2}+O(\eta^{-6}).
\label{eq:H_farfield}
\end{equation}
The first correction, $-2\eta^{-2}$, is universal, whereas the coefficient $C$ is determined by the global solution. The leading-order behavior $H(\eta)\sim C\eta^2$ is equivalent to requiring that
\begin{equation}
q(\xi,\tau)=|\tau_*-\tau|^{1/2}
H\left(\frac{\xi}{|\tau_*-\tau|^{1/4}}\right)
\end{equation}
becomes independent of $\tau$ sufficiently far from the collapsing core.

To formulate the fourth-order problem numerically on a finite interval $0\leq\eta\leq L$, we impose symmetry at the local minimum at $\eta=0$,
\begin{equation}
H'(0)=0,\qquad H'''(0)=0,
\end{equation}
and impose the asymptotic behavior \eqref{eq:H_farfield}, up to terms including $O(\eta^{-2})$, at $\eta=L$. Eliminating the unknown coefficient $C$ using
\begin{equation}
H''(L)=2C-\frac{12}{L^4}
\end{equation}
gives the two outer boundary conditions
\begin{equation}
H(L)=\frac{L^2}{2}H''(L)+\frac{4}{L^2},
\qquad
H'(L)=LH''(L)+\frac{16}{L^3}.
\label{eq:H_outer_BCs}
\end{equation}
The far-field coefficient is subsequently recovered from the numerical solution according to
\begin{equation}
C=\frac12\left[H''(L)+\frac{12}{L^4}\right].
\label{eq:C_extract}
\end{equation}

An exact integral relation provides a useful consistency check on the numerical profiles. Integrating Eq.~\eqref{eq:nlbvp} from $0$ to an arbitrary $R$ and using the symmetry conditions at the origin gives
\begin{equation}
H'''(R)+\frac{H'(R)}{H(R)}
+\frac{RH(R)}{4}
-\frac34\int_0^R H(\eta)\,d\eta=0.
\label{eq:H_integral_identity}
\end{equation}
This relation holds for every $R$ for any sufficiently regular symmetric solution of Eq.~\eqref{eq:nlbvp} and provides a stringent check involving both the local behavior and the integral of the numerical profile. Interestingly, taking the infinite-domain limit, $R\to \infty$, and using the asymptotic expansion \eqref{eq:asymp_exp}, Eq.~\eqref{eq:H_integral_identity} implies the simple exact identity
\begin{equation}
\int_0^\infty\left[H(\eta)-C\eta^2\right] d\eta
= 0. \label{eq:exact_relation_intH}
\end{equation}

\begin{figure}
    \centering
    \includegraphics[width=\columnwidth]{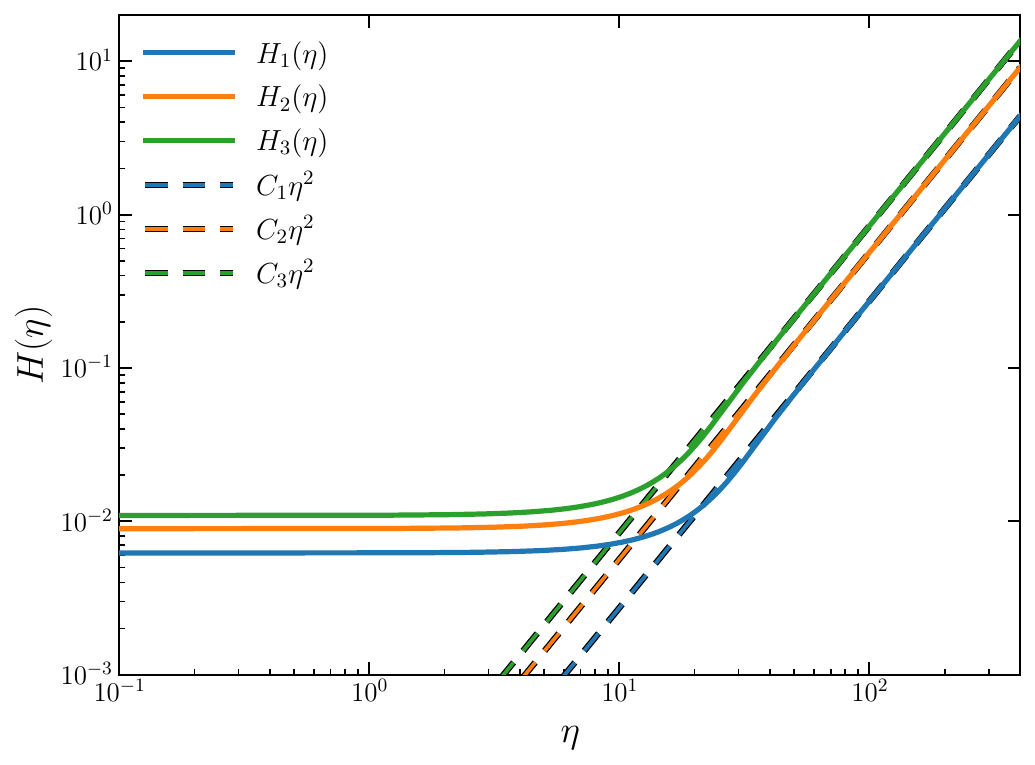}
    \caption{Three distinct positive similarity profiles $H(\eta)$ obtained by solving Eq.~\eqref{eq:nlbvp} from the initial guesses $H_{\rm guess}(\eta)=0.01+C_0\eta^2$ with $C_0=10^{-5}$, $2\times10^{-5}$, and $3\times10^{-5}$. The corresponding far-field coefficients are $C\simeq2.79\times10^{-5}$, $5.68\times10^{-5}$, and $8.45\times10^{-5}$, respectively. Dashed curves show the leading far-field behavior $H(\eta)\sim C\eta^2$.}
    \label{fig:nlbvp_selfsimilar}
\end{figure}

We solve the nonlinear fourth-order boundary value problem formulated above using the adaptive collocation method implemented in SciPy's \textit{solve\_bvp} routine \cite{2020SciPy-NMeth}. 
Our numerical solutions indicate that the boundary value problem  does not have a  unique solution. To illustrate this multiplicity, we solve the problem from three independently specified initial guesses of the form
\begin{equation}
H_{\rm guess}(\eta)=0.01+C_0\eta^2,
\end{equation}
with $C_0=10^{-5}$, $2\times10^{-5}$, and $3\times10^{-5}$. For $L=400$, a maximum of $300,000$ collocation nodes, and a solver tolerance of $10^{-8}$, these three initializations converge to three distinct positive, monotonically increasing profiles, shown in Fig.~\ref{fig:nlbvp_selfsimilar}. The corresponding far-field coefficients are
\begin{equation}
C\simeq
2.79\times10^{-5},\quad
5.68\times10^{-5},\quad
8.45\times10^{-5},
\label{eq:C_three_profiles}
\end{equation}
with
\begin{equation}
H(0)\simeq
0.00628,\quad 0.00896,\quad 0.01094,
\end{equation}
respectively. In all three cases $H''(0)>0$ and the minimum of the computed profile occurs at $\eta=0$.

The three independently initialized BVP calculations therefore converge to distinct positive similarity profiles. Each profile satisfies the exact finite-domain integral identity \eqref{eq:H_integral_identity} to near machine precision, providing a stringent numerical consistency check beyond convergence of the collocation solver itself. We have also verified that the numerical profiles approximately satisfy Eq.~\eqref{eq:exact_relation_intH}, where we used the asymptotic expansion to represent the solution beyond $\eta=L$, further corroborating the numerically obtained solutions. It is important to stress that the three solutions for $H(\eta)$ shown in Fig.~\ref{fig:nlbvp_selfsimilar} are not exhaustive, they are simply illustrative elements of the solution space of the boundary value problem.

The coexistence of these distinct numerical profiles is reminiscent of the multiplicity found in the thin-film rupture problem of Zhang and Lister \cite{ZhangLister1999}. Indeed, the coexistence of multiple self-similar solutions is well known in a wide range of applications \cite{witelski1999stability,eggers2008role,bernoff2010stability}. A more detailed classification of the solution set of Eq.~\eqref{eq:nlbvp}, including whether the admissible profiles form a discrete family or belong to one or several continuous branches, requires a more detailed analysis of the nonlinear boundary value problem and is left for future work.

If a phase slip occurs at $\xi=\xi_*$, the similarity scaling predicts that, as $\tau\to\tau_*$, the pattern amplitude behaves as
\begin{align}
R
&=\sqrt{1-k^2}
\approx\sqrt{2q}
\nonumber\\
&=\sqrt{2}\,|\tau_*-\tau|^{1/4}
\sqrt{
H\left(
\frac{|\xi-\xi_*|}
{|\tau_*-\tau|^{1/4}}
\right)
}.
\label{eq:similarity_ansatz}
\end{align}
The appropriate similarity profile must ultimately be determined by matching to the time-dependent solution. It is important to reiterate that, as a consequence of the long-wave limit used in deriving the WKB equation, the reduced dynamics does not describe the physics at the phase slip itself.

The self-similar analysis assumes $R,k\gg\epsilon$ and therefore remains valid only as long as $ \tau_*-\tau\gg\epsilon^4.$ Consequently, the self-similar description breaks down when $\tau_*-\tau=O(\epsilon^4)$, suggesting a temporal boundary layer of this scale and a corresponding spatial core of width $O(\epsilon)$, within which the reduced equation is no longer applicable. We emphasize that this conclusion relies on the hyperdiffusive regularization considered here, for which the
hyperdiffusivity is assumed, for simplicity, to be state-independent. If the coefficient $\kappa$ itself becomes singular as the phase slip is approached (see Appendix~\ref{app:hi_ord_wkb}), the asymptotic structure may change qualitatively. We leave this
problem for future study.

Finally, in addition to \textit{defect chaos} involving numerous phase slips, locally described by the analysis above, the unscaled  CGLE also possesses a regime known as \textit{phase chaos}, in which the amplitude fluctuates near $R=1$ and no phase slips occur \cite{aranson2002world}. We have found no evidence of such behavior in solutions of Eq.~(\ref{eq:nl_diff_eq_full}).

\subsection{Validation against direct numerical simulations}
\label{sec:validation_ss_dns}

To verify the above theoretical predictions for the self-similar behavior of solutions near phase slips, we resort to DNS of Eq.~\eqref{eq:nl_diff_eq_full} using a Fourier pseudospectral solver with periodic boundary conditions. Time integration is performed using a fourth-order exponential time-differencing Runge--Kutta scheme (ETDRK4) \cite{kassam2005fourth}. As the phase slip is approached and the characteristic spatial scale of the solution progressively decreases, the Fourier resolution is successively increased. Refinement is controlled by the width $W_{1.5}$, defined as the distance between two crossings of $q=1.5q_{\min}$, where $q_{\min}$ is the minimum value of $q$ at a given time. For a given spatial resolution $\Delta \xi$, we check whether $W_{1.5}/\Delta\xi$ decreases to below $14$ grid points, in which case the solution is spectrally interpolated onto a grid with twice the number of grid points and Fourier modes. Data from successive resolutions are retained over an overlap region extending to $W_{1.5}/\Delta\xi\simeq 22$, allowing consistency between consecutive refinement levels to be assessed. This procedure is repeated using up to $N=2^{20}$ Fourier modes; at the highest resolution, the calculation is finally continued until $W_{1.5}/\Delta\xi\simeq6$.
\begin{figure}
\centering

\begin{overpic}[width=\linewidth]{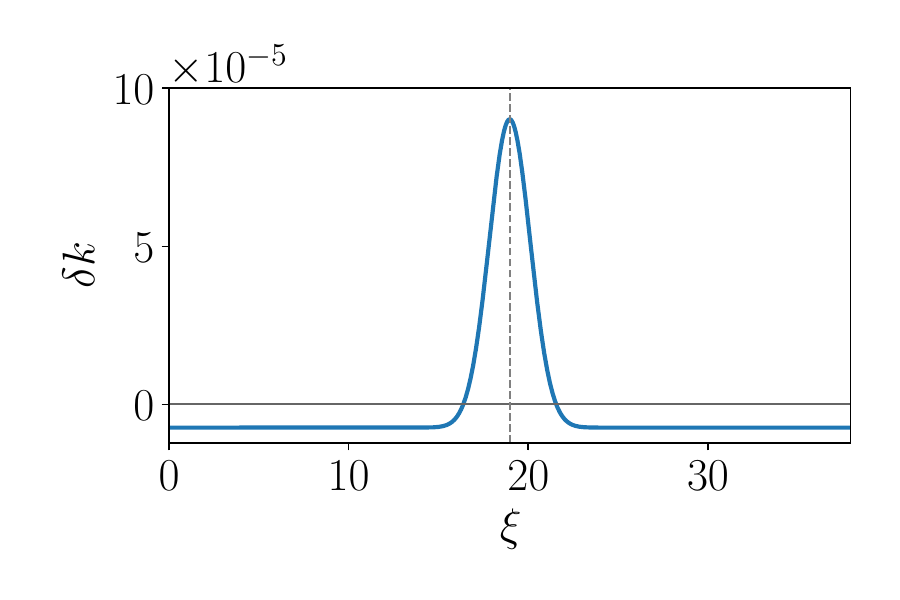}
\put(3,60){\color{black} (a)}
\end{overpic}

\vspace{-0.2em}

\begin{overpic}[width=\linewidth]{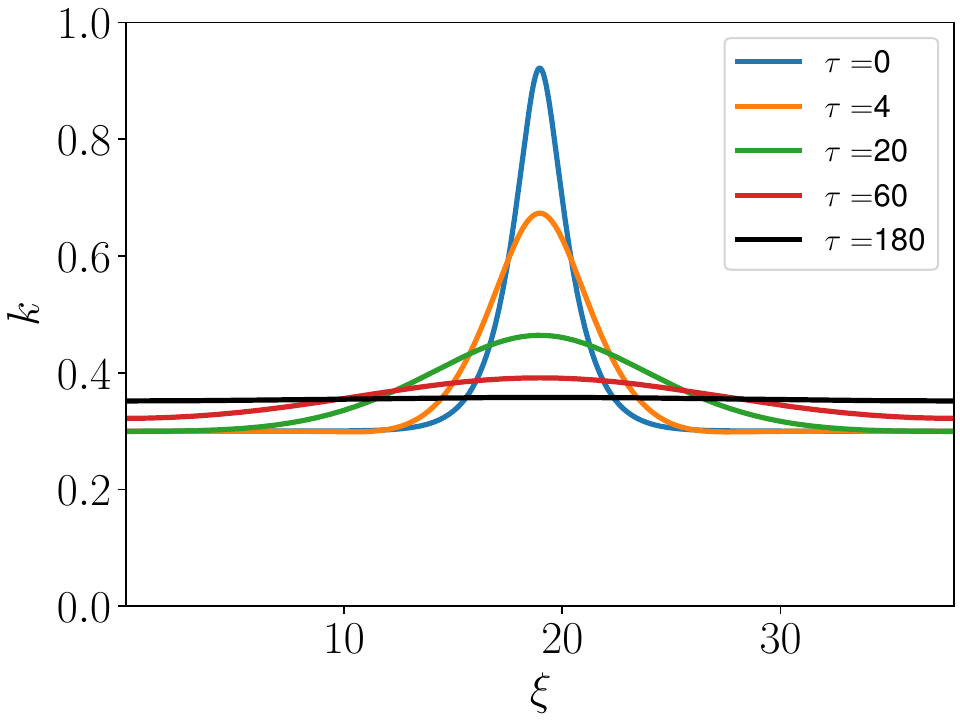}
\put(-2,72){\color{black} (b)}
\end{overpic}

\vspace{-0.1em}

\begin{overpic}[width=\linewidth]{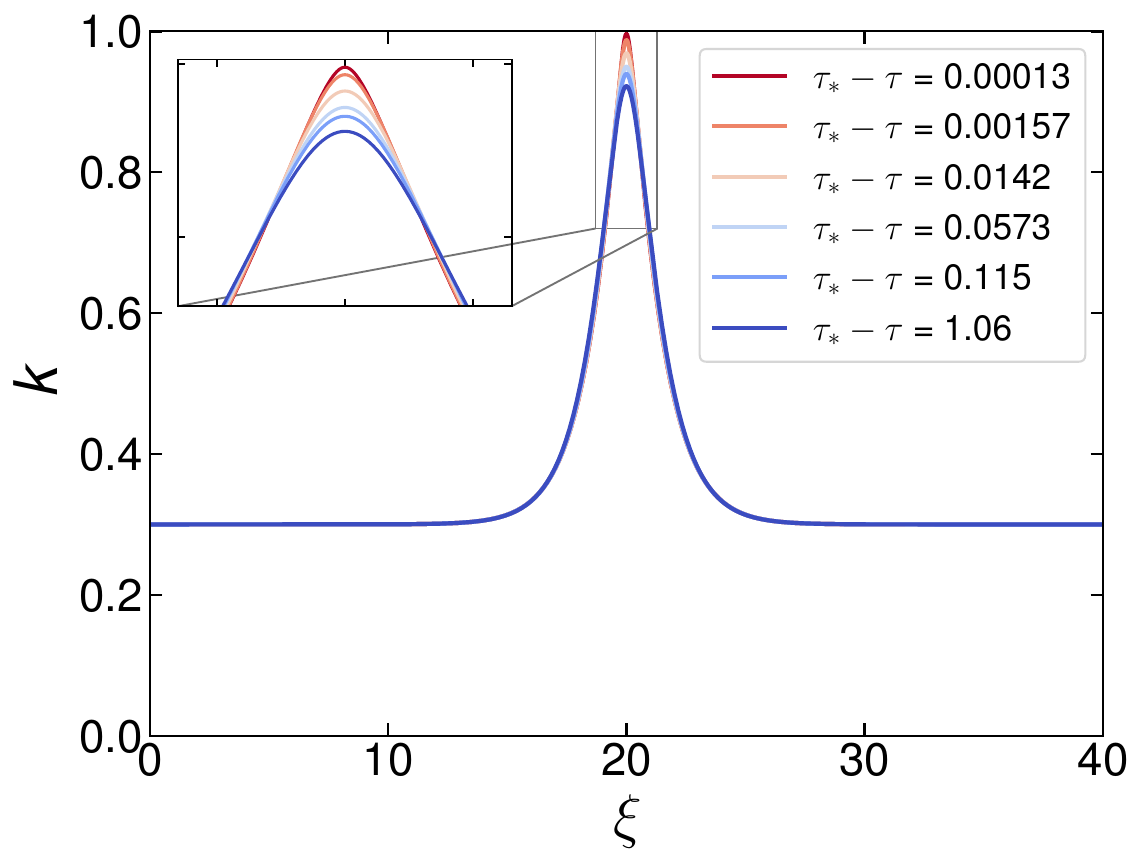}
\put(0,67){\color{black} (c)}
\end{overpic}

\caption{The homoclinic pulse acts as a separatrix between decay to a uniform state and phase-slip dynamics. (a) Localized perturbation added to or subtracted from the steady pulse asymptoting to $k=0.3$. (b) Subtraction leads to decay toward a uniform state. On an infinite domain $k(\xi\to\infty)$ is fixed, while in the finite periodic domain considered here conservation of $\int k\,d\xi$ forces $k$ away from the peak to increase as the peak itself decays. (c) Addition triggers a finite-time phase slip, at which the diffusion coefficient in Eq.~\eqref{eq:nl_diff_eq_full} becomes singular.}
\label{fig:growth_and_decay_single_pulse}
\end{figure}

Specifically, we consider the case $\alpha=\beta$, so that the Burgers term vanishes from Eq.~\eqref{eq:nl_diff_eq_full}. We initialize the simulation with the unstable homoclinic pulse state with asymptotic wave number $k=0.3$ shown in Fig.~\ref{fig:pulse}, to which a small-amplitude, mean-zero perturbation is applied. Figure~\ref{fig:growth_and_decay_single_pulse}(a) shows the perturbation profile, which features a peak at the location of the homoclinic peak profile (not shown) and changes sign away from the peak, ensuring that the domain average of the perturbation vanishes. Despite the small amplitude of $O(10^{-4})$ of the perturbation, it is found to make a significant difference in the evolution of the perturbed steady solution: the unstable steady state, which corresponds to a saddle point of the free energy, separates trajectories leading to a phase slip from those resulting in diffusive relaxation toward a uniform state.

In Fig.~\ref{fig:growth_and_decay_single_pulse}(b), the initial perturbation in panel (a) is subtracted from the homoclinic pulse. The pulse broadens diffusively while decaying in amplitude, and the final state of this evolution is one with uniform $k$ given by the conserved spatial average. Figure~\ref{fig:growth_and_decay_single_pulse}(c) shows the temporal evolution from the initial peak when the perturbation in panel (a) is added rather than subtracted. The maximum grows toward $k=1$, i.e., a coefficient singularity, in finite time. In the latter case, we can compare the numerical evolution against the predicted self-similar dynamics near the singularity derived above.

\begin{figure}
    \centering
    \includegraphics[width=\linewidth]{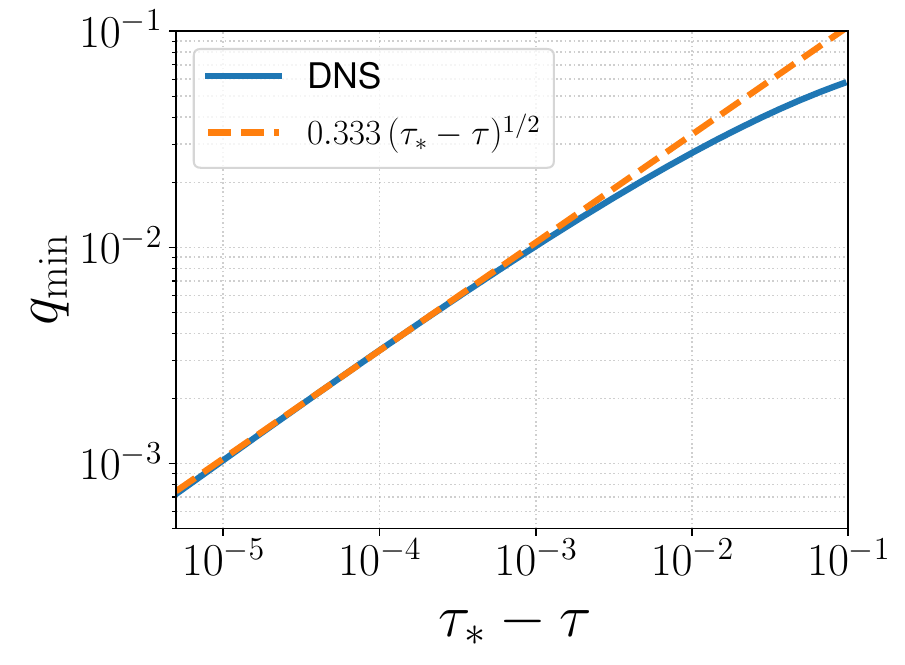}

    \caption{Minimum amplitude $q_{\min}$ as a function of the time remaining until the phase slip, $\tau_*-\tau$. The numerical solution closely follows the square-root scaling $q_{\min}\propto(\tau_*-\tau)^{1/2}$ predicted by the self-similar solution.}
    \label{fig:qmin_vs_time}
\end{figure}

To probe the amplitude scaling, we plot in Fig.~\ref{fig:qmin_vs_time} the minimum value $q_{\min}$, where $q\equiv1-k$, as the solution approaches the phase slip at $k=1$. The similarity solution predicts
\begin{equation}
q_{\min}\propto(\tau_*-\tau)^{1/2}.
\end{equation}
The numerical results closely follow this square-root scaling over several decades, consistent with the temporal scaling predicted by the similarity solution.

The similarity solution also predicts the evolution of the spatial scale of the singular region. In particular, the characteristic width $W$ (to be defined) is predicted to shrink proportionally with $(\tau_*-\tau)^{1/4}$, which, together with $q_{\min}\propto(\tau_*-\tau)^{1/2}$, gives the equivalent prediction
\begin{equation}
W\propto q_{\min}^{1/2}.
\end{equation}
To test this spatial scaling directly without relying on the fitted singularity time $\tau_*$, we consider $W_2$, defined as the distance between the two points at which $q=2q_{\min}$.

\begin{figure}
    \centering
    \includegraphics[width=0.9\linewidth]{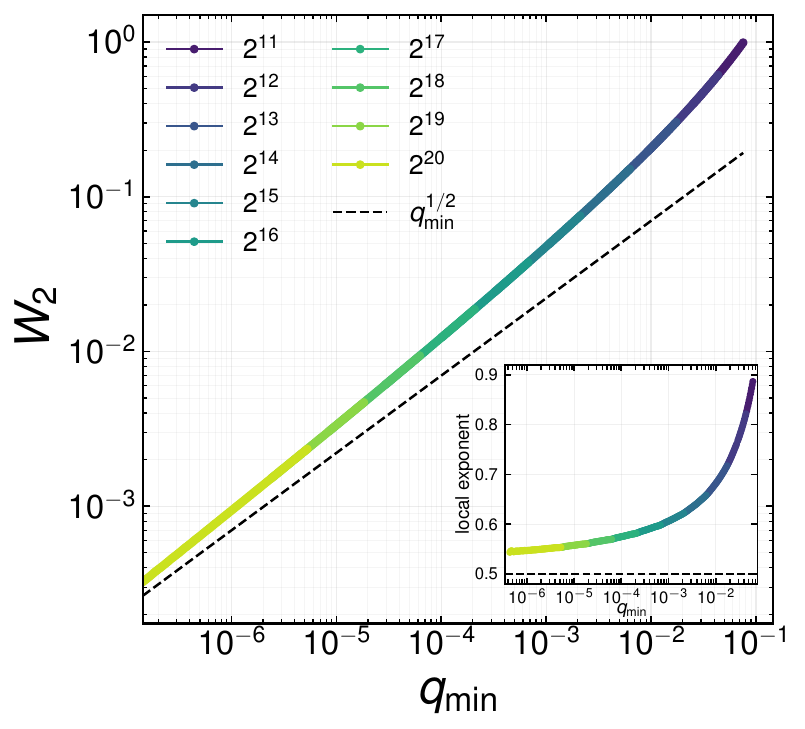}

    \caption{Width $W_2$, defined by the two crossings of $q=2q_{\min}$, as a function of $q_{\min}$, with colors indicating successive spatial resolutions $N=2^{11},\ldots,2^{20}$. The dashed line shows the predicted scaling $W_2\propto q_{\min}^{1/2}$. The inset shows the corresponding local exponent $d\ln W_2/d\ln q_{\min}$, which slowly approaches the theoretical value $1/2$ as $q_{\min}\to0$, reaching $\lesssim0.55$ at the smallest resolved $q_{\min}$. 
    }
    \label{fig:W2_vs_qmin}
\end{figure}

Figure~\ref{fig:W2_vs_qmin} shows $W_2$ as a function of $q_{\min}$ over successive spatial refinements. The different colors indicate the resolutions used as the singular region shrinks, from $N=2^{11}$ to $N=2^{20}$. As the phase slip is approached, the numerical solution progressively approaches the predicted scaling $W_2\propto q_{\min}^{1/2}$. The local logarithmic exponent $d\ln W_2/d\ln q_{\min}$, shown in the inset, decreases toward the theoretical value $1/2$ and reaches $\lesssim0.55$ at the smallest resolved $q_{\min}$. The slow approach indicates an extended prescaling regime before the asymptotic similarity limit is reached.

\section{Complex-coefficient GLE}
\label{sec:cgle}
In this section, we extend the diffusion equation approximation to the CGLE, where at least one of $\alpha\neq \beta$ is nonzero. In this case \eqref{eq:nl_diff_eq_full} is still reflection-symmetric but the reflection is now ${\cal R}: (\xi,k)\to -(\xi,k)$. The presence of the symmetry ${\cal R}$ is important since it permits the existence of solutions satisfying $k(\xi)=-k(-\xi)$. This symmetry is broken by solutions $k=k_0\ne0$.

\subsection{Dispersion relation}
\textcolor{black}{
The WKB dispersion relation~\eqref{eq:disp_rel} generalizes to the complex-coefficient case:
 \begin{align}
     \sigma(q) = \underbrace{-D(k_0)q^2 - \kappa q^4 }_{\equiv \sigma_r}+\underbrace{2i\left(\beta-\alpha\right) k_0 q }_{\equiv -i\omega}.
 \end{align} 
 The real part $\sigma_r$ of the growth rate describes diffusive decay or antidiffusive growth, while $\omega=2(\alpha-\beta)k_0 q$ corresponds to the frequency of small-amplitude perturbations, implying that the group velocity $c_g=d\omega/dq=2(\alpha-\beta)k_0$.
}
\subsection{Exact shock solution\label{sec:front_CGLE}}
 For $\kappa=0$, i.e., at leading order in the WKB approximation, we can give an exact traveling front solution. Consider a traveling solution with a velocity $c$, and let $z=\xi-c\tau$. Then
 \begin{align}
 0=&\frac{\mathrm{d}}{\mathrm{d}z}\left(\frac{1-3k^2}{1-k^2}k^{\prime}\right)+\left[2\left(\beta-\alpha\right) k+c\right]k^{\prime} \\ 
 \Rightarrow A=&\frac{1-3k^2}{1-k^2}k^{\prime}+\left[\left(\beta-\alpha\right) k+c\right]k.\end{align}
The constant $A$ is in general nonzero, except in the special case where $k$ vanishes at either $x\to\pm \infty$, i.e., the case of a front connecting a spatially uniform state and a patterned state. Further integration yields
 \begin{align}z-z_0=\int\frac{1-3k^2}{1-k^2}\frac{1}{A-\left[\left(\beta-\alpha\right) k+c\right]k}dk
\label{eq:front_profile_integral_z_of_k},
 \end{align}
where the integral is known analytically, see Appendix~\ref{app:exact_front}. We are interested in front solutions, which connect two roots $k_1,k_2$ of 
\begin{equation}
    k^{\prime}(k)=\frac{A-[\left(\beta-\alpha\right) k+c]k}{D(k)} = \frac{(\alpha-\beta)(k-k_1)(k-k_2)}{D(k)},
\end{equation}
where $1/D(k)=(1-k^2)/(1-3k^2)$ is nonzero, and the speed $c$ of the front and the constant flux $-A$ are fully determined by $k_1,k_2$ given $\alpha,\beta$,
\begin{equation} c = (\alpha-\beta)(k_1+k_2),\quad A = (\alpha-\beta)k_1k_2. \label{eq:c_and_A_for_front}
\end{equation}

The result~\eqref{eq:c_and_A_for_front} for the speed $c$ is identical to the Rankine-Hugoniot condition across the front and agrees with the known propagation speed of hole solutions in the cubic Ginzburg-Landau equation at arbitrary $\epsilon$ obtained by matching the asymptotic phase speeds on both sides of the front; see Eq.~(37) of \cite{aranson2002world}.

\begin{figure}
    \centering
    \includegraphics[width=1.05\linewidth]{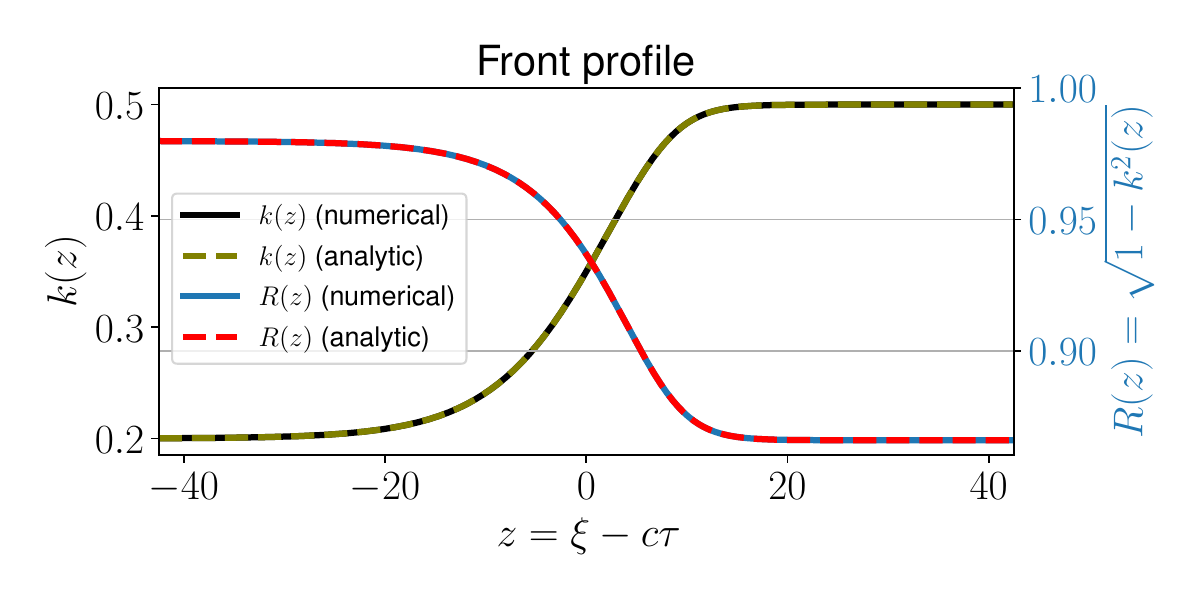}
    \caption{Front solution of the second-order nonlinear diffusion equation in terms of $k(z)$ and $R(z)=\sqrt{1-k(z)^2}$ with $\alpha=1$, $\beta=3/2$, connecting two asymptotic plane waves with wave numbers $k_1=0.2$ and $k_2=0.5$. Dashed lines indicate the exact solution given in Appendix~\ref{app:exact_front}.}
    \label{fig:front_second_order}
\end{figure}

Clearly, $D(k)$ cannot vanish between $k_1,k_2$ for a well-defined front to exist or else the slope becomes singular, i.e., fronts exist only between two wave numbers in the Eckhaus-stable or Eckhaus-unstable regimes, respectively, but cannot cross the Eckhaus boundary $k=1/\sqrt{3}$. For $k_1,k_2\in (0,1/\sqrt{3})$ and $\alpha>\beta$, $k'(k)$ is negative between the two roots, and is positive between them when $\beta>\alpha$ (see Fig.~\ref{fig:front_second_order}). It can easily be verified that the group velocity on either side of the front is directed into the shock in the comoving frame, i.e., the shock is a wave sink as expected. Related wave-interaction phenomena, including nonstationary oscillatory shocks, were studied by Malomed \cite{malomed1983nonsteady}. In contrast, for $1/\sqrt{3}<k_1,k_2<1$ (similar to the fronts whose existence is rigorously proved for the real-coefficient GLE in \cite{eckmann1993front}), one finds that $k'(k)$ is positive for $\alpha>\beta$, while it is negative for $\beta>\alpha$. The resulting (unstable) front is a wave \textit{source} rather than a sink, indicating that it should be referred to as a hole rather than a shock, in keeping with the established CGLE nomenclature~\cite{aranson2002world,sandstede2004defects}.

Aranson and Kramer \cite{aranson2002world} write that \textit{there are no exact analytic expressions for the sink solution connecting two traveling waves with arbitrary wave numbers}. This observation is consistent with the more recent exhaustive classification of meromorphic traveling-wave solutions by Conte et al. \cite{conte2023all}, which contains no such sink solution for the cubic CGLE. In particular, the classical Nozaki-Bekki solutions \cite{nozaki1984exact,BekkiNozaki1985} do not include any sink solution that connects two plane waves with distinct wave numbers and nonzero amplitudes. In contrast, the solution presented in Eq.~\eqref{eq:front_profile_integral_z_of_k} provides just such an expression, connecting any two wave numbers exclusively in the Eckhaus-stable (or unstable) regime, valid in the small-$\epsilon$ limit of Eq.~(\ref{eq:cgl}). The fact that our exact solution of the reduced equation is valid for two \textit{arbitrary} asymptotic wave numbers (in the Eckhaus-stable or -unstable bands) implies that it cannot be meaningfully compared with any of the known exact solutions of the full CGLE, such as the classical Nozaki-Bekki solutions.

\subsection{Comparison with  numerical solutions of the complex Ginzburg-Landau equation}
\label{sec:comp}
Here, we present simulations of the full CGLE and compare the results with the theoretical predictions from WKB theory detailed above. 

In the simulations, we represent the complex order parameter field $u(\xi,\tau)$ by its local amplitude $R(\xi,\tau)$ and phase $\phi(\xi,\tau)$ as $u(\xi,\tau)=R(\xi,\tau)\exp(i\phi/\epsilon)$, as in the WKB ansatz, but without performing an expansion in $\epsilon$. Rewriting Eq.~(\ref{eq:cgl}) in terms of these variables yields 
\begin{align}
\epsilon^{2} R_{\tau}
=& R + \epsilon^{2} R_{\xi\xi}
- (\phi_{\xi})^{2} R - R^{3}  \label{eq:cgl_R}\\& \nonumber
- 2 \epsilon^{2} \alpha\, R_{\xi} \phi_{\xi}
- \epsilon^{2} \alpha\, R \phi_{\xi\xi},
\\
R \,\phi_{\tau}
=& \epsilon^{2} \alpha\, R_{\xi\xi}
- \alpha\, R (\phi_{\xi})^{2}
+ 2 R_{\xi} \phi_{\xi}  \label{eq:cgl_theta}\\
&+ R \phi_{\xi\xi}
+ \beta R^{3}. \nonumber
\end{align}
Dividing Eq.~(\ref{eq:cgl_theta}) by $R$ (assumed to be nonzero) and differentiating with respect to $\xi$, we find the following equation for the wave number $k=\phi_\xi$, 
\begin{align}
k_{\tau}
=& \partial_{\xi} \left[
  - \alpha k^{2}
  + 2 \frac{R_{\xi}}{R}\,k
  + k_{\xi}
  + \beta R^{2}
\right]\notag \\&
+ \epsilon^{2} \partial_{\xi} \left[
  \alpha\,\frac{R_{\xi\xi}}{R}
\right].\label{eq:cgl_k}
\end{align}
We have implemented a sparse numerical solver in C++ for Eqs.~(\ref{eq:cgl_R}) and (\ref{eq:cgl_theta}), the latter divided by $R$, using a second-order finite-difference scheme for all spatial derivatives and a first-order implicit-explicit Euler time marching scheme, wherein all linear terms are treated implicitly and all nonlinear terms explicitly. Specifically, in Eq.~(\ref{eq:cgl_R}), $R+\epsilon^2R_{\xi\xi}$ is treated implicitly, while in Eq.~(\ref{eq:cgl_theta}), only $\theta_{\xi\xi}$ is treated implicitly. As boundary conditions, we impose a constant wave number $\phi_\xi = k_1$ at the left end of the domain and $\phi_\xi = k_2$ at the right end, as well as the corresponding amplitudes $R=\sqrt{1-k_1^2}$ on the left and $R=\sqrt{1-k_2^2}$ on the right. 

\subsubsection{Spatial eigenvalues}
A well-known feature \cite{aranson2002world} of the complex GLE, Eq.~(\ref{eq:cgl}), is that in two spatial dimensions it has spatially localized spiral wave solutions with exponentially decaying tails. Importantly, interactions between such spiral waves, when they are spatially separated, are mediated by these exponential tails and have been studied extensively \cite{Rica1990interaction,aranson1991interaction_PhysD,aranson1991interaction_PRL,rica1992dynamics,aranson1993theory,aranson2002world}. In particular, it is known \cite{aranson1993theory} that there is a critical boundary in the $(\alpha,\beta)$ plane separating monotonic, exponentially decaying tails from oscillatory, exponentially decaying tails of spiral waves. Oscillatory spiral wave tails are found for $ (\beta-\alpha)/(1+\beta \alpha) > C\approx 0.845
$ and monotonic spiral wave tails otherwise. These results may be rephrased in the language of spatial dynamics, which has been widely used \cite{kirchgassner1982wave,haragus2010local} to analyze the structure and interactions of spatially localized states \cite{knobloch2015spatial,kirchgassner1982wave,haragus2010local,burke2008classification,burke2012localized,parra2018bifurcation,knobloch2021stationary,parra2021origin,verschueren2021dissecting,frohoff2021localized,al2021localized,li2025traveling}, including in the Swift-Hohenberg equation \cite{burke2006localized,raja2023collisions} and in the Lugiato-Lefever equation \cite{parra2018bifurcation}. Monotonic tails correspond to real spatial eigenvalues, while oscillatory tails reflect complex spatial eigenvalues. Specifically for interactions between spatially localized structures, knowing whether the structures in question have monotonic or oscillatory tails is important, since bound states exist generically when the two overlapping tails are both oscillatory (corresponding to complex spatial eigenvalues), as established theoretically and numerically in many systems \cite{gorshkov1981interactions,aranson1990stable,vladimirov2002two,tlidi2003interaction,tlidi2008vegetation,clerc2010interaction}. 

Here, we perform a similar analysis of the spatial eigenvalues associated with front solutions connecting two traveling waves with wave numbers $k\to k_1$ as $\xi\to -\infty$, $k\to k_2$ as $\xi\to\infty$. We show that a similar transition takes place between a monotonic and an oscillatory regime, characterized by real and complex spatial eigenvalues, respectively. The regime boundary in the present case is different from that for spiral waves due to the important dependence on the propagation velocity $c=(\alpha-\beta)(k_1+k_2)$ of the traveling front solutions, which explicitly involves the asymptotic wave numbers $k_1$ and $k_2$, in addition to $\alpha, \beta$.

Linearizing Eqs.~(\ref{eq:cgl_R}), (\ref{eq:cgl_k}) about a traveling wave with a given constant wave number $k=k_0$ and associated amplitude $R=R_0\equiv\sqrt{1-k_0^2}$ (where $k_0$ will be taken to be either $k_1$ or $k_2$), and writing $R=R_0+r$, $k=k_0+v$, we find
\begin{align}
\epsilon^{2} r_{\tau}\label{eq:cgl_lin_r}
&= \epsilon^{2} r_{\xi\xi}
   - 2 R_{0}^{2}\,r
   - 2 k_{0} R_{0}\,v
    \\
&\quad
   - 2\epsilon^{2}\alpha k_{0}\, r_{\xi}
   - \epsilon^{2}\alpha R_{0}\,v_{\xi},\nonumber\\
v_{\tau}\label{eq:cgl_lin_k}
&= v_{\xi\xi}
   + \frac{2k_{0}}{R_{0}}\, r_{\xi\xi}
   + \epsilon^{2}\alpha\,\frac{r_{\xi\xi\xi}}{R_{0}}
   \\
&\quad
   - 2\alpha k_{0}\,v_{\xi}
   - 2\beta R_{0}\, r_{\xi}. \nonumber
\end{align}

We seek a solution that is stationary in a frame translating at velocity $c=(\alpha-\beta)(k_1+k_2)$. Letting $z=\xi-c\tau $, and using the ansatz $(r(z),v(z))=(\hat{r},\hat{v})e^{\lambda z}$ with constant $\hat{r},\hat{v}$, one obtains the characteristic polynomial for the spatial eigenvalue $\lambda$
\begin{align}
\lambda \; (A\,\lambda^{3}
\;+\;
B\,\lambda^{2}
\;+\;
C\,\lambda
\;+\;
D)
= 0,
\label{eq:char_poly}
\end{align}
where
\begin{align}
\begin{aligned}
A =& \alpha^{2}\epsilon^{4} + \epsilon^{2},\\
B =& 2(\alpha-\beta)(k_1+k_2)\,\epsilon^{2},\\
C =& 6k_{0}^{2} - 2+\epsilon^{2} \left[ \bigl(c - 2\alpha k_{0}\bigr)^{2}- 2\alpha\beta\,(1 - k_{0}^{2})\right],\\
D =& -2(1-k_{0}^{2})\,[c - 2k_{0}(\alpha-\beta)].
\end{aligned}
\end{align}

Clearly, $\lambda=0$ is always a root of Eq.~(\ref{eq:char_poly}); this is a consequence of the invariance of the system under $\phi\to\phi+{\rm const.}$ Generically, one of the remaining roots, that with real part closest to zero, controls the shape of the linear tail of the front, as we shall discuss further below.

Two distinct slices through the parameter space $(\epsilon, \alpha, \beta)$ will be considered:
\begin{itemize}
    \item[a.] First, we vary $\alpha,\beta$ for $\epsilon=1$ (unscaled CGLE). In the limit $\alpha,\beta\to 0$, one simply recovers the RGLE.
    \item[b.] Next, we fix $\alpha,\beta$ and study how the spatial eigenvalue spectrum depends on $\epsilon$. In the limit $\epsilon\to 0$, which not only sends the imaginary parts of the coefficients in the CGLE to zero but also simultaneously takes the long-wave limit, one formally recovers the WKB approximation discussed in previous sections.
\end{itemize}

\begin{figure}
    \centering
    \includegraphics[width=1.05\linewidth]{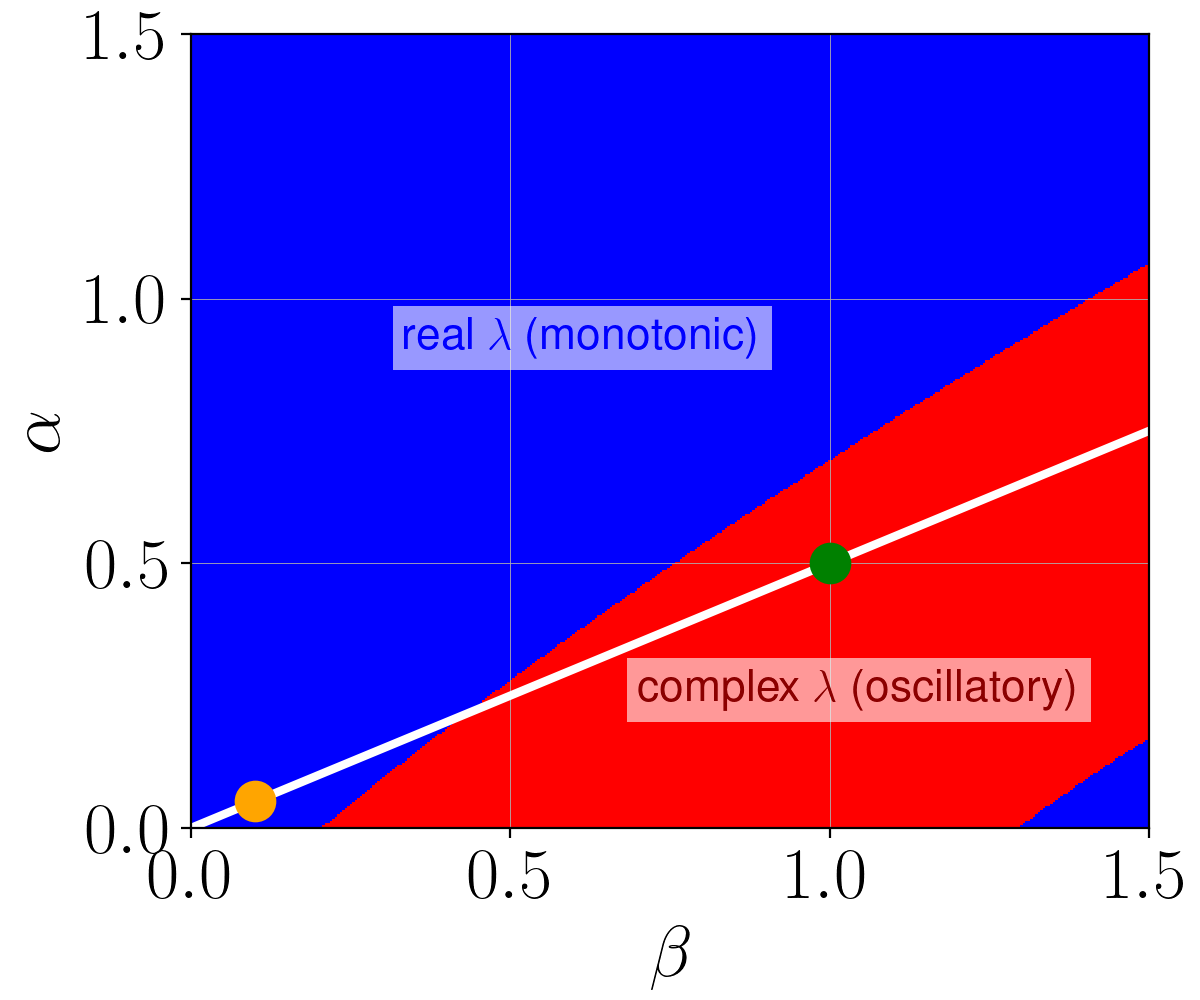}
    \caption{Regime diagram (deduced from the solutions of Eq.~(\ref{eq:char_poly})) shown in terms of $\alpha,\beta$ for $\epsilon=1$, $k_1=0.2$, $k_2=0.5$, with the linearization being about $k_0= k_2=0.5$ (the tail towards $\xi\to \infty$). The blue and red areas are defined in terms of the dominant decaying spatial eigenvalue, i.e., that with negative real part that is closest to zero. Blue areas indicate the regions in parameter space where said dominant spatial eigenvalue is real (monotonic exponential tail), while red areas indicate where it is complex (oscillatory tail). The white line indicates the path $\alpha=\beta/2$ through the parameter space for which the full spatial eigenvalue spectrum is shown in Fig.~\ref{fig:spatial_eps1_correct}. The orange and green circles indicate reference parameter values for which we compare our DNS results with the spatial eigenvalue predictions (Fig.~\ref{fig:comparison profiles}).} \label{fig:regime_diagram_spat_ev_alpha_beta_eps1}
\end{figure}

\paragraph{Varying $\alpha,\beta$ at fixed $\epsilon$:}
We fix $\epsilon=1$ (unscaled GLE) and vary $\alpha,\beta$, analyzing the roots of Eq.~(\ref{eq:char_poly}). To determine the shape of the exponential tail of the front connecting the two asymptotic wave numbers $k_1,k_2$, it is essential to identify the dominant eigenvalue, i.e., that whose real part is closest to zero. For $\xi \to \infty$, the relevant eigenvalue $\lambda$ must have negative real part, while the real part must be positive for the tail at $\xi\to - \infty$. Figure~(\ref{fig:regime_diagram_spat_ev_alpha_beta_eps1}) shows the regime diagram in the $(\alpha,\beta)$ plane based on the linearization about $k=k_2= 0.5$. Blue areas indicate where in parameter space the dominant spatial eigenvalue at $\xi\to\infty$ is real (front with monotonic tail at $\xi\to\infty$), while red areas indicate where it is complex (front with oscillatory tail at $\xi\to\infty$). Performing the same analysis for $\xi \to -\infty$, where $k_0=k_1=0.2$, one finds that the dominant eigenvalue is always real for the values of $(\alpha,\beta,\epsilon)$ used in Fig.~\ref{fig:regime_diagram_spat_ev_alpha_beta_eps1} (not shown). This indicates that for this choice of asymptotic wave numbers, and within this region of parameter space, only the tail towards $\xi\to\infty$ may change from monotonic to oscillatory, while the tail towards $\xi\to-\infty$ is always monotonic. We stress that the regime diagram shown in Fig.~\ref{fig:regime_diagram_spat_ev_alpha_beta_eps1} is not universal, but depends sensitively on the choice of $k_0$. This is in stark contrast to two-dimensional spiral-wave solutions, where there is a simple curve in the $(\alpha,\beta)$ plane that forms the regime boundary separating oscillatory from monotonic tails~\cite{aranson1993theory}. Indeed,
it is not uncommon to find one oscillatory and one monotonic tail in systems with propagating localized structures, e.g., \cite{kawahara1988pulse,cross1993pattern,kalliadasis2007thin,li2025traveling}.

\begin{figure}
    \centering  \includegraphics[width=1.1\linewidth]{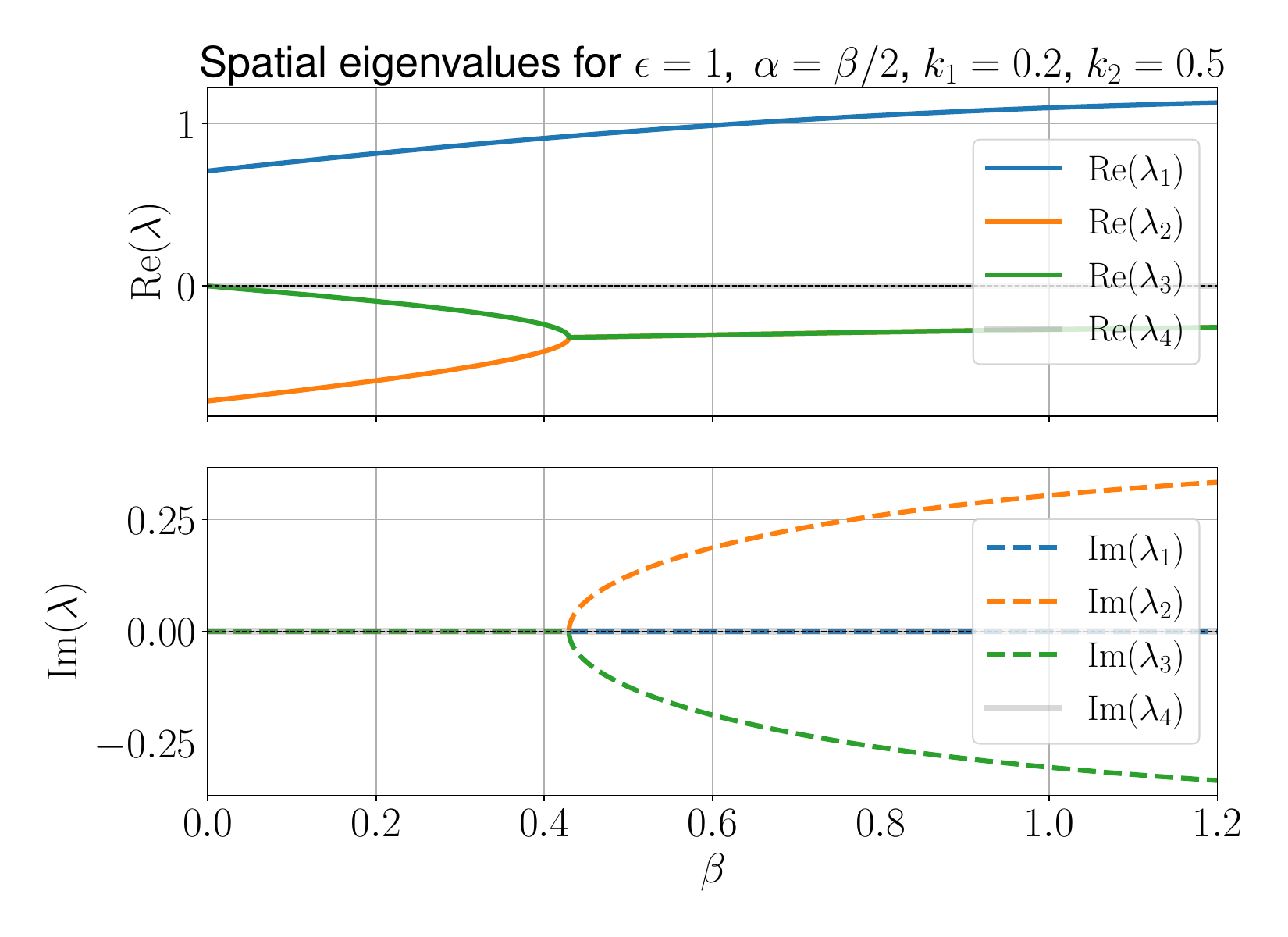}
    \label{fig:spatial_eps1}
    \caption{Spectrum of spatial eigenvalues (top panel: real part, bottom panel: imaginary part) for $\epsilon=1,\, \alpha=\beta/2,\, k_1=0.2,\,k_2=0.5$, where the linearization is performed about a traveling wave with wave number $k_2=0.5$. The transition from a real to complex dominant eigenvalue occurs at $\beta\approx0.42$, in agreement with Fig.~\ref{fig:regime_diagram_spat_ev_alpha_beta_eps1}. 
    }
    \label{fig:spatial_eps1_correct}
\end{figure}

\begin{figure}
    \centering
    \includegraphics[width=1.05\linewidth]{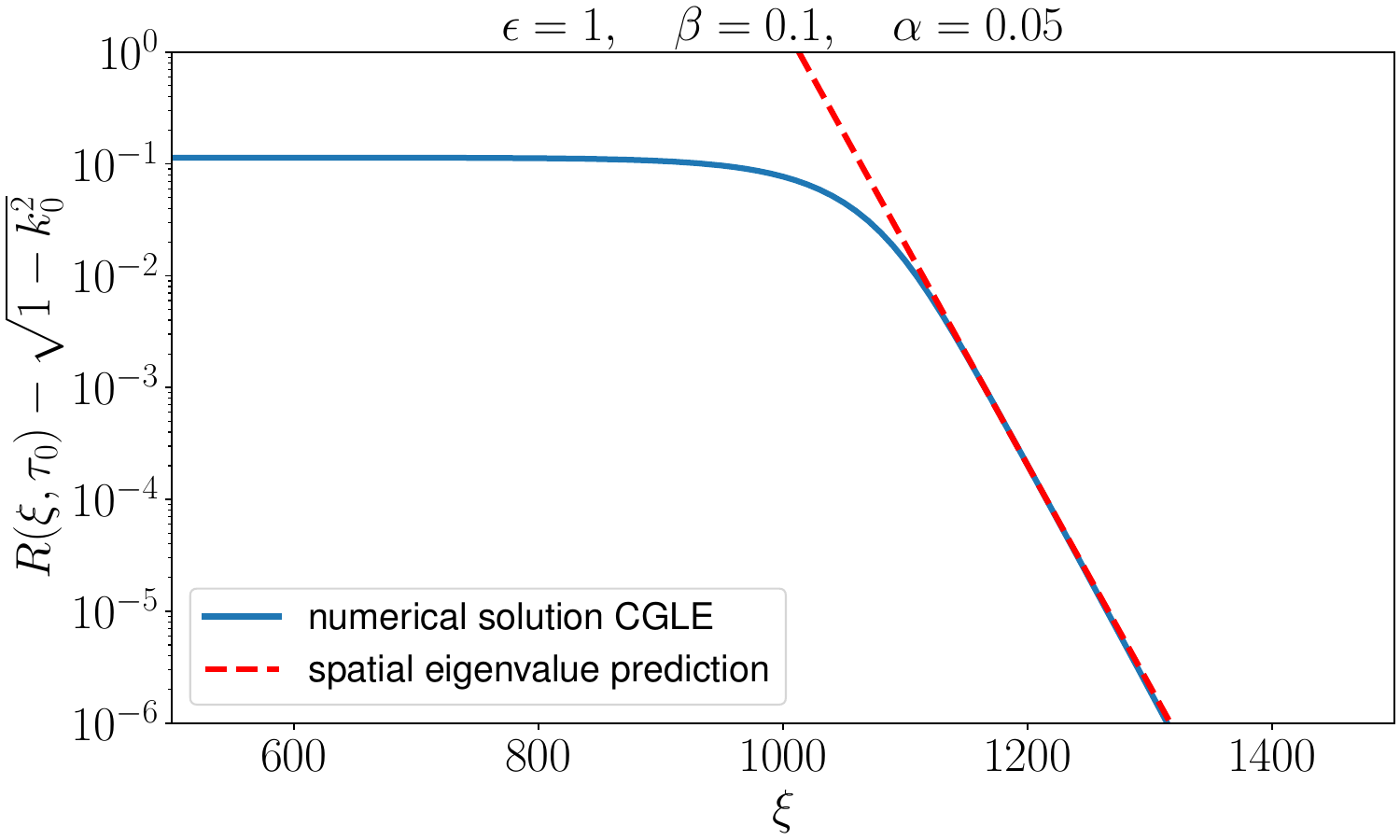}\\ \includegraphics[width=1.05\linewidth]{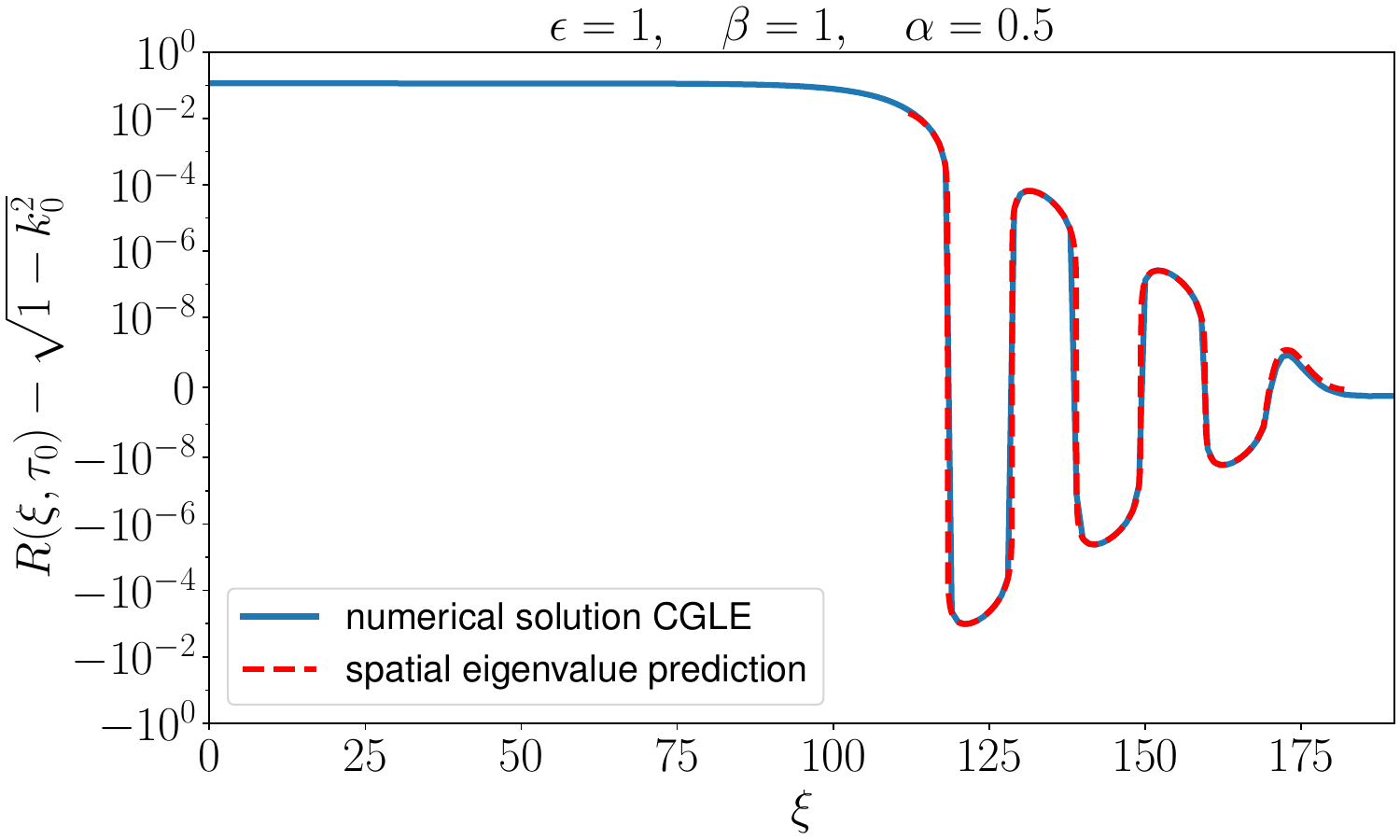} 
    \caption{Front solutions from numerical integration of the CGLE versus spatial eigenvalue predictions for $\epsilon=1$ (unrescaled CGLE). The simulation results are shown at a late time $t_0$, where they are steady in the comoving frame, in terms of the spatial profile of the deviation (on a logarithmic scale) of the amplitude $R$ from $\sqrt{1-k_0^2}$, the asymptotic value of $R$ at $\xi\to\infty$. Top: at $\beta=0.1$, $\alpha=0.05$ (orange dot in Fig.~\ref{fig:regime_diagram_spat_ev_alpha_beta_eps1}) the front connects \textit{monotonically} $k_1=0.2$ at $\xi\to -\infty$ to $k_2=0.5$ at $\xi\to +\infty$. The decay scale of the exponential agrees with the spatial eigenvalue prediction (dominant eigenvalue $\lambda\approx  -0.0454$) shown by the red dashed line. Bottom: at $\beta=1$, $\alpha=0.5$ (green dot in Fig.~\ref{fig:regime_diagram_spat_ev_alpha_beta_eps1}), the tail is oscillatory and both decay rate and wavelength agree with the spatial eigenvalue prediction (dominant eigenvalue $\lambda\approx-0.288 + 0.242i $) shown by the red dashed curve. }
    \label{fig:comparison profiles}
\end{figure}

Figure~\ref{fig:spatial_eps1_correct} shows the full spatial eigenvalue spectrum (all four roots of Eq.~(\ref{eq:char_poly})) along the path through parameter space indicated by the white line in Fig.~\ref{fig:regime_diagram_spat_ev_alpha_beta_eps1}, i.e., for $\epsilon=1$ and $\beta=2\alpha$. At $\beta\approx 0.43$, the dominant spatial eigenvalue (which has negative real part) switches from real to complex. Note that for $\alpha=\beta=0$, one recovers the spatial eigenvalues of the RGLE, namely, $\lambda=0$ (double root due to phase translation invariance and an additional conserved gradient structure) and $\lambda=\pm \sqrt{2-6k_0^2}/\epsilon$~($=\pm 1/\sqrt{2}$ for $k_0=0.5$, $\epsilon=1$ as assumed here, in agreement with Fig.~\ref{fig:spatial_eps1_correct}).

Figure~\ref{fig:comparison profiles} shows two front solutions of the CGLE obtained at late time using time integration, one with a monotonic tail and one with an oscillatory tail. These correspond to the orange and green dots in Fig.~\ref{fig:regime_diagram_spat_ev_alpha_beta_eps1}, respectively. The red dashed lines in both cases indicate the corresponding tail structure predicted by the spatial eigenvalue calculation, which agrees well with the numerically obtained profile not only in terms of the exponential decay rate, but also the wavelength of the oscillating tail in the complex eigenvalue case. In both cases, the monotonic decay of the tail as $\xi\to-\infty$ is consistent with the real spatial eigenvalue found in this limit.

\begin{figure}
    \centering
    \includegraphics[width=\linewidth]{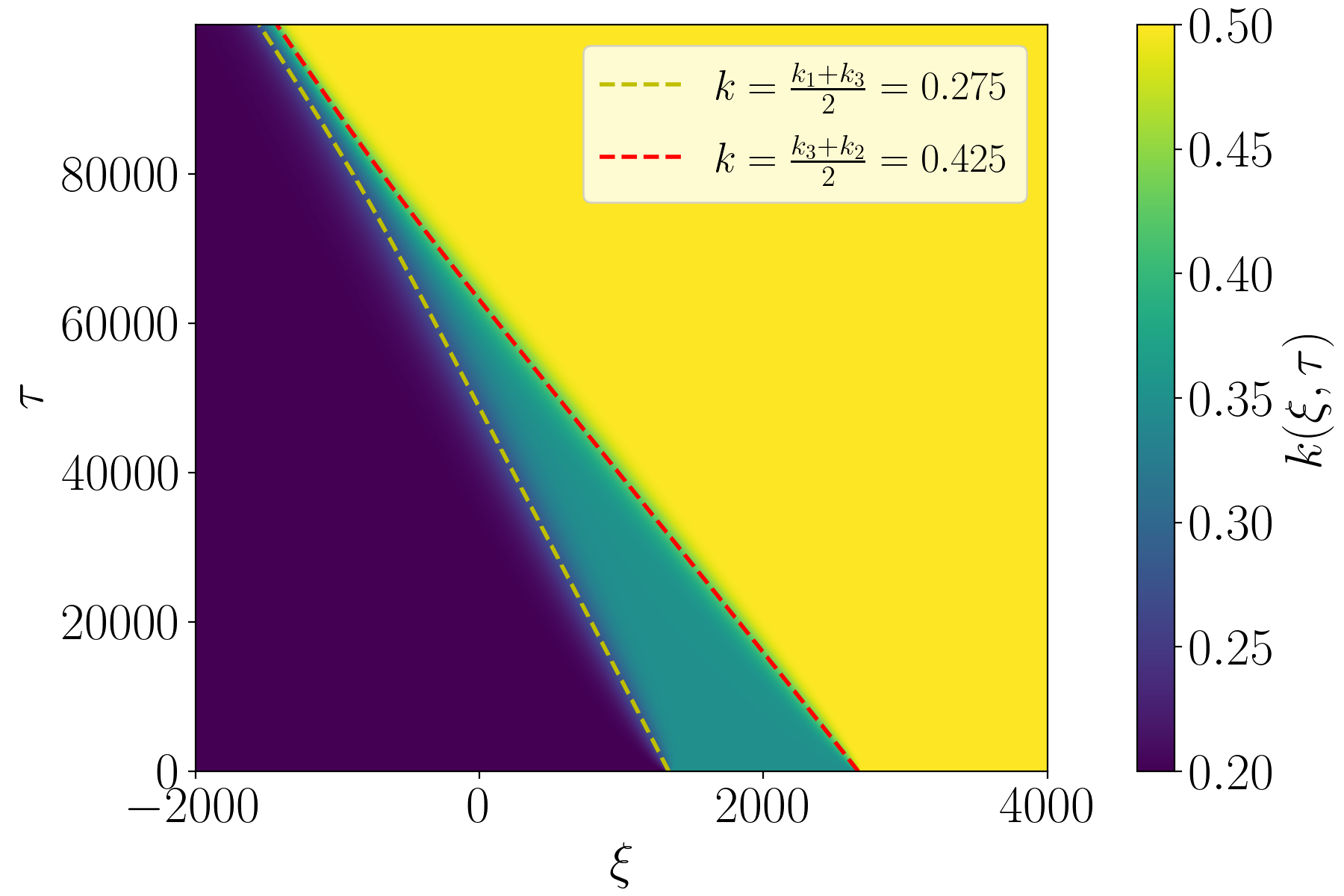}
    \caption{Space-time diagram of a collision between traveling waves with asymptotic wave numbers $k_1=0.2$, $k_2=0.5$ and an intermediate wave number $k_{3}=0.35$, for $\epsilon=1$, $\beta=0.1$, $\alpha=0.05$ (orange dot in Fig.~\ref{fig:regime_diagram_spat_ev_alpha_beta_eps1}). The two fronts merge into a single front separating domains with $k_1$ and $k_2$ since no bound state at finite separation can form, given that both overlapping tails are monotonic rather than oscillatory. }
    \label{fig:spacetime_plot_collision}
\end{figure}

As mentioned earlier, knowledge of the spatial eigenvalues associated with localized structures in a given system is useful since it allows one to derive qualitative and quantitative understanding about the interactions between the structures. Figure~\ref{fig:spacetime_plot_collision} shows a space-time diagram obtained by integrating the CGLE with $\beta=0.1$, $\alpha=0.05$, $\epsilon=1$ (same parameter values as the orange dot in Fig.~\ref{fig:regime_diagram_spat_ev_alpha_beta_eps1}) from an initial condition with two smoothed-out steps in the wave number: $k=k_1=0.2$ in the leftmost third of the domain, $k=k_3=0.35$ in the middle third and $k=k_2=0.5$ in the rightmost third of the domain. The resulting fronts both travel in the negative $\xi$ direction, since $c=(\alpha-\beta)(k_1+k_2)<0$. The rightmost front propagates faster than the leftmost front and therefore eventually catches up with it. Since the overlapping tails are monotonic (not shown), no bound state at a finite separation can exist and therefore the two fronts merge into a single front from $k_1$ to $k_2$. We have performed a stochastic search in terms of $k_1$, $k_2$, $\alpha$, $\beta$ and $\epsilon$ to identify whether there are any regions in parameter space where the two overlapping tails are both oscillatory, but have not found any such configuration. Although this does not constitute a rigorous proof, it suggests that collisions between two fronts in the CGLE consistently result in a single front, with no possibility of forming a bound state at finite separation.

\paragraph{Varying $\epsilon$ at fixed $\alpha, \beta$:}

\begin{figure}
    \centering
    \includegraphics[width=\linewidth]{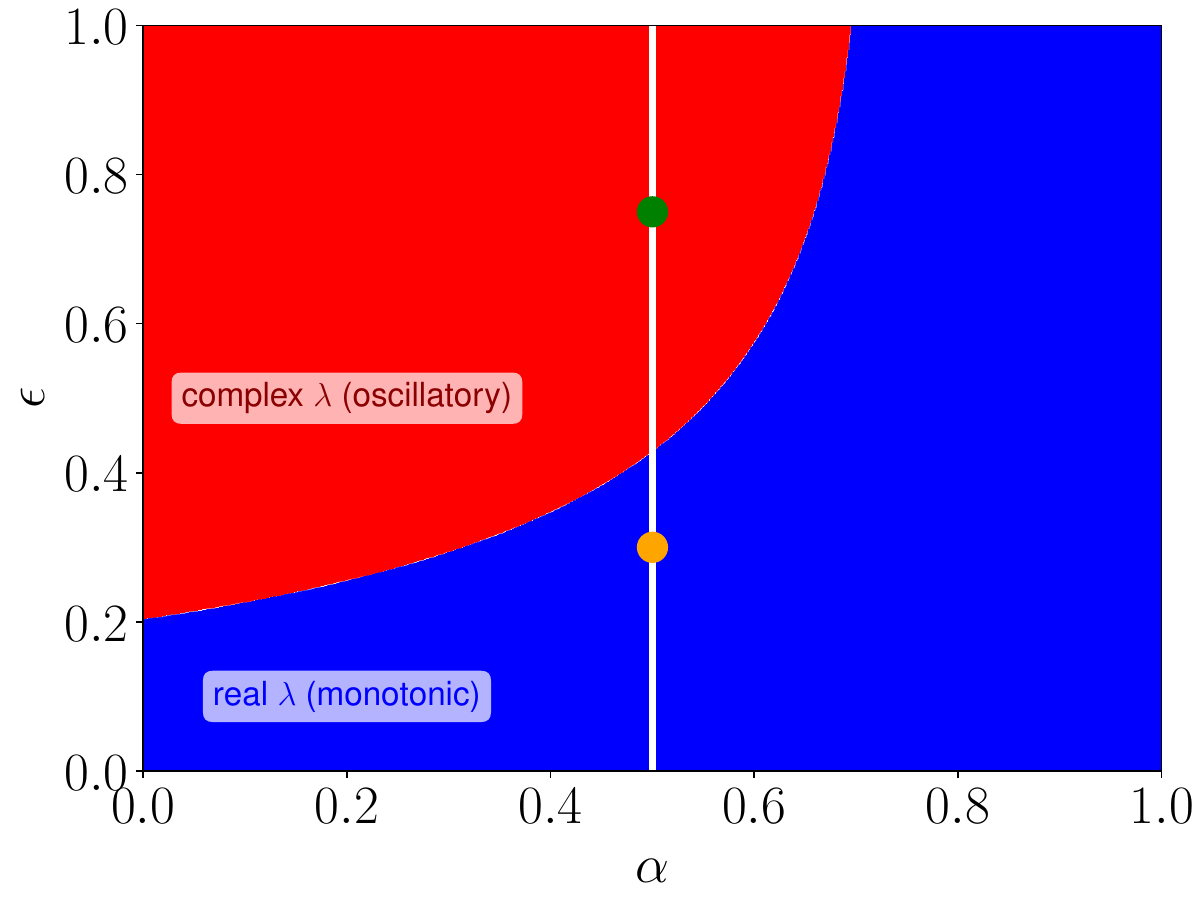}
    \caption{Regime diagram for $\beta=1$ and varying $\epsilon,\alpha$ when $k_1=0.2$, $k_2=k_0=0.5$. Blue and red areas indicate regions where the dominant eigenvalue with negative real part (controlling the approach to $k=k_2$ at $\xi\to \infty$) is real or complex, respectively. The WKB limit corresponds to the limit $\epsilon\to 0$ from above. 
    The white vertical line corresponds to the path through parameter space shown in Fig.~\ref{fig:spatial_eigenvalues_vs_eps}.}
    \label{fig:regime_diagram_alpha_eps_beta1}
\end{figure}

For $\alpha,\beta$ fixed, one finds that in the distinguished limit $\epsilon\to 0$ two roots of Eq.~(\ref{eq:char_poly}) diverge at the rate $O(1/\epsilon)$, since the limit $\epsilon\to 0$ is singular. In addition to the root $\lambda=0$ which invariably remains, the only finite, nonzero spatial eigenvalue reads
\begin{equation}
    \lambda = \frac{(\alpha-\beta)[2k_0-(k_1+k_2)](1-k_0^2)}{
    1-3 k_0^2}.
    \label{eq:spat_eval_eps0_analyical}
\end{equation}
This result can also be derived directly by linearizing the nonlinear wave number diffusion equation~(\ref{eq:nl_diffusion_equation_complex_2nd_order}) about a traveling wave with wave number $k=k_0$ (either $k_1$ or $k_2$) in the frame moving at $c=(\alpha-\beta)(k_1+k_2)$, where $k_1,k_2$ are the asymptotic wave numbers as $\xi \to\pm \infty$. Hence, in the limit $\epsilon \to 0$, the dominant spatial eigenvalue is invariably real. This theoretical result is consistent with the regime diagram for different $\alpha$ and $\epsilon$ at fixed $\beta=1$ shown in Fig.~\ref{fig:regime_diagram_alpha_eps_beta1}. As in Fig.~\ref{fig:regime_diagram_spat_ev_alpha_beta_eps1}, red color indicates the region where the leading spatial eigenvalue about $k_0=k_2=0.5$ (with $k_1=0.2$) is complex, while blue color indicates where it is real. Figure~\ref{fig:regime_diagram_alpha_eps_beta1} confirms that, for sufficiently small values of $\epsilon$, the dominant eigenvalue is indeed real. 

\begin{figure}
    \centering
    \includegraphics[width=1.075\linewidth]{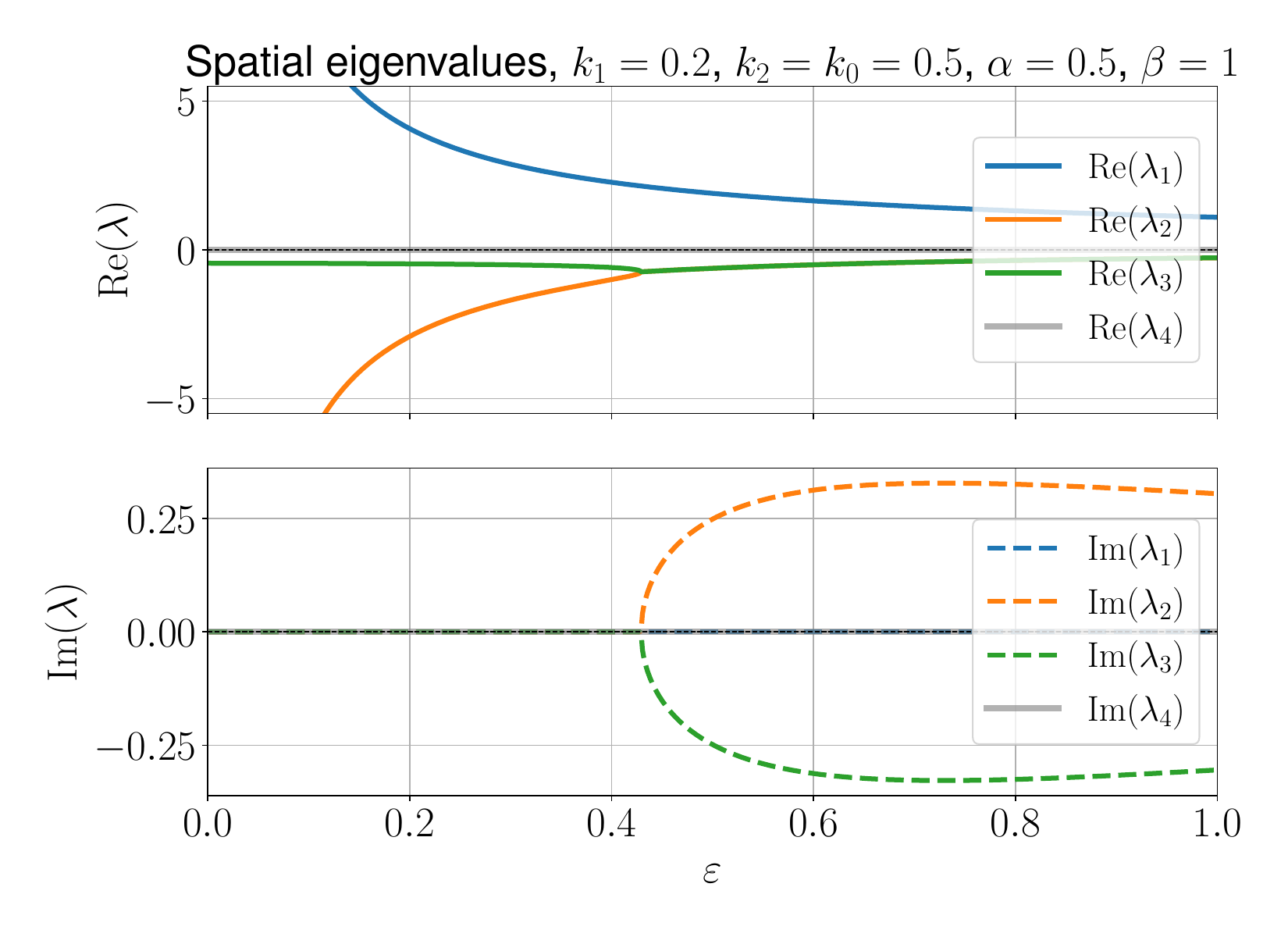}
    \caption{Spatial eigenvalues $\lambda$ as functions of $\epsilon$ at fixed $\alpha=0.5$, $\beta=1$ for asymptotic wave numbers $k_1=0.2$, $k_2=0.5$ from linearization about a traveling wave with wave number $k_0=k_2$ (tail toward $\xi\to\infty$). At $\epsilon\approx 0.43$ the dominant spatial eigenvalue transitions from real to complex. Note that only two finite eigenvalues remain in the limit $\epsilon\to 0$, in agreement with Eq.~(\ref{eq:nl_diffusion_equation_complex_2nd_order}).}
    \label{fig:spatial_eigenvalues_vs_eps}
\end{figure}

\begin{figure}
    \centering
    \includegraphics[width=\linewidth]{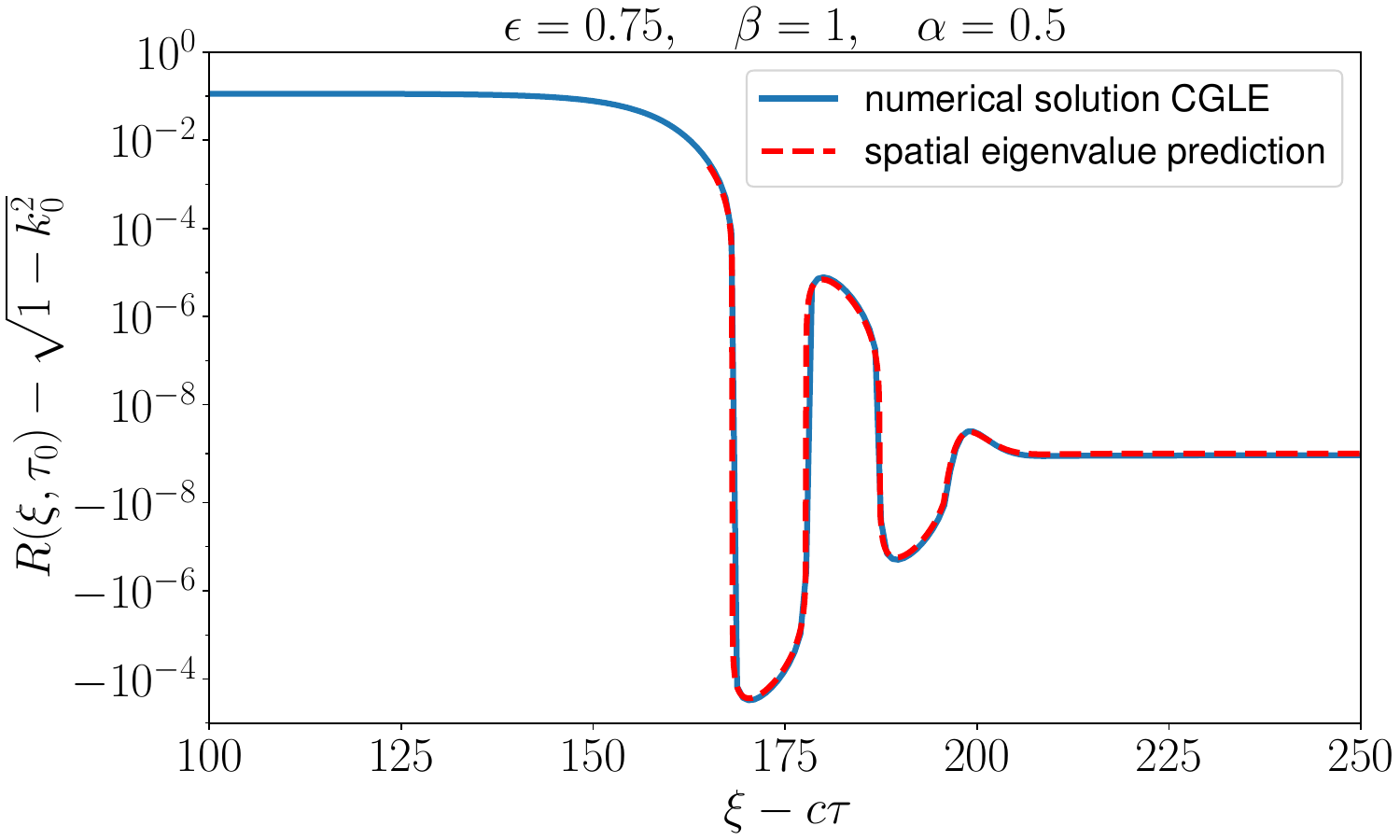}
        \includegraphics[width=\linewidth]{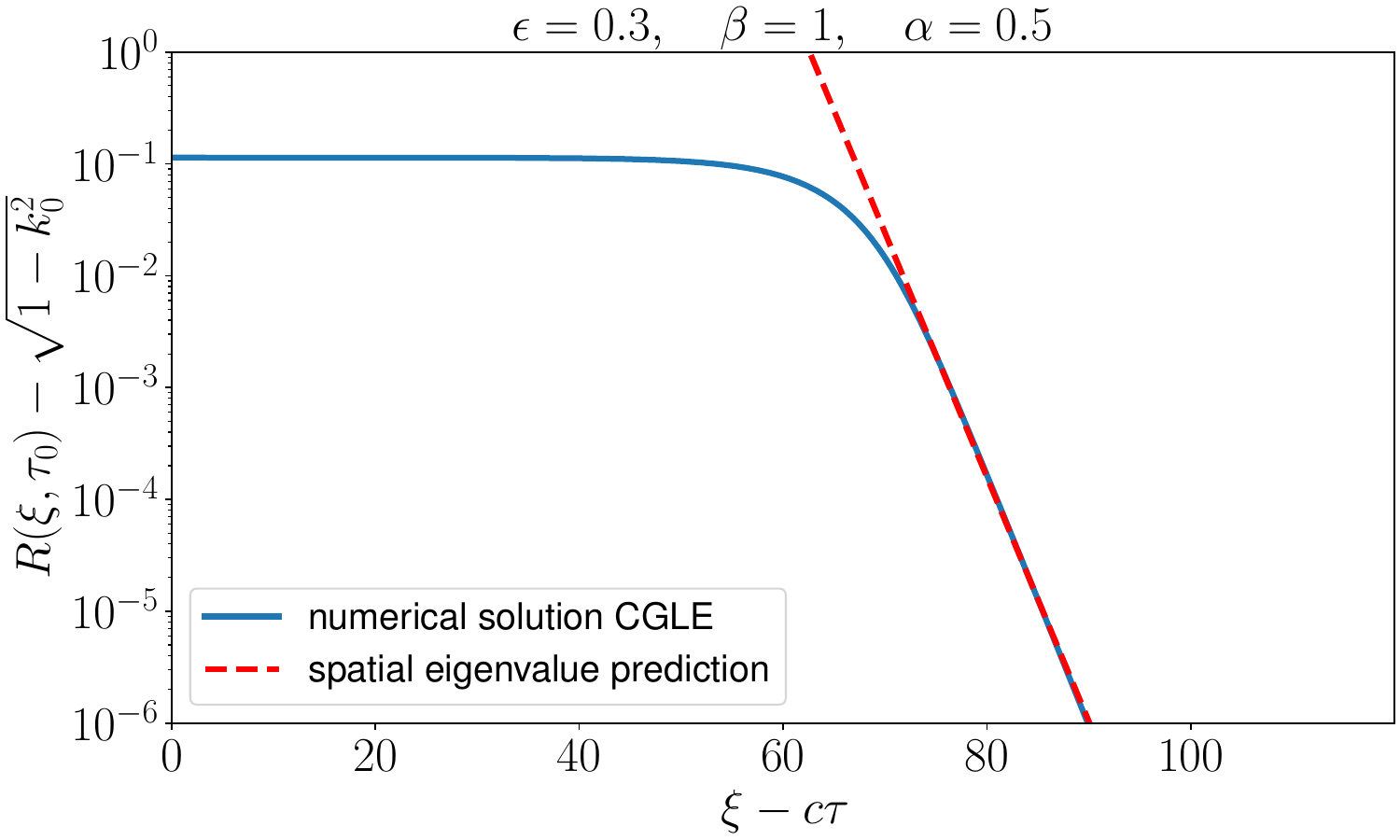}
    \caption{Front solutions from numerical integration of the CGLE versus spatial eigenvalue predictions, for $\alpha=0.5$, $\beta=1$ and two values of $\epsilon$, similar to Fig.~\ref{fig:comparison profiles}. Top panel: $\epsilon=0.75$, bottom panel: $\epsilon=0.3$, corresponding to the green and orange dots in Fig.~\ref{fig:regime_diagram_alpha_eps_beta1}, respectively. The red dashed line indicates the predicted tail profile based on the dominant spatial eigenvalue, and agrees well with the numerical results not only for the sp
    atial decay rate but also for the wavelength in the oscillatory tail. In the top panel, $\lambda\approx  -0.383158+0.327578i$, while in the bottom panel $\lambda\approx -0.50275$.}
    \label{fig:solution_profiles_at_eps0.75_and_eps0.3_vs_spat_eval}
\end{figure}

Figure~\ref{fig:spatial_eigenvalues_vs_eps} shows all four roots of Eq.~(\ref{eq:char_poly}) for the linearization about $k_0=k_2=0.5$ (as before $k_1=0.2$) as functions of $\epsilon$ for $\beta=1$, $\alpha=0.5$, i.e., along the path indicated by the white vertical line in Fig.~\ref{fig:regime_diagram_alpha_eps_beta1}. We note the anticipated divergence of two of the eigenvalues as $O(1/\epsilon)$, as well as the convergence of the sole nonzero eigenvalue to a negative constant in the limit $\epsilon\to 0$, whose value ($-0.45$ for the values chosen here) agrees with Eq.~(\ref{eq:spat_eval_eps0_analyical}). Figures~\ref{fig:regime_diagram_alpha_eps_beta1} and \ref{fig:spatial_eigenvalues_vs_eps} consistently indicate that the transition from oscillatory to monotonic tails occurs slightly above $\epsilon= 0.4$. Figure~\ref{fig:solution_profiles_at_eps0.75_and_eps0.3_vs_spat_eval} shows two solutions of the CGLE obtained by time integration, one at $\epsilon=0.75$ and one at $\epsilon=0.3$, corresponding to the green and orange dots in Fig.~\ref{fig:regime_diagram_alpha_eps_beta1}, respectively. At $\epsilon=0.75$ (top panel of Fig.~\ref{fig:solution_profiles_at_eps0.75_and_eps0.3_vs_spat_eval}) the tail is oscillatory, while at $\epsilon=0.3$, it is monotonic. In both cases, the predicted tail structure based on the dominant spatial eigenvalue agrees well with the numerically obtained CGLE solutions.

\begin{figure}
    \centering
    \includegraphics[width=\linewidth]{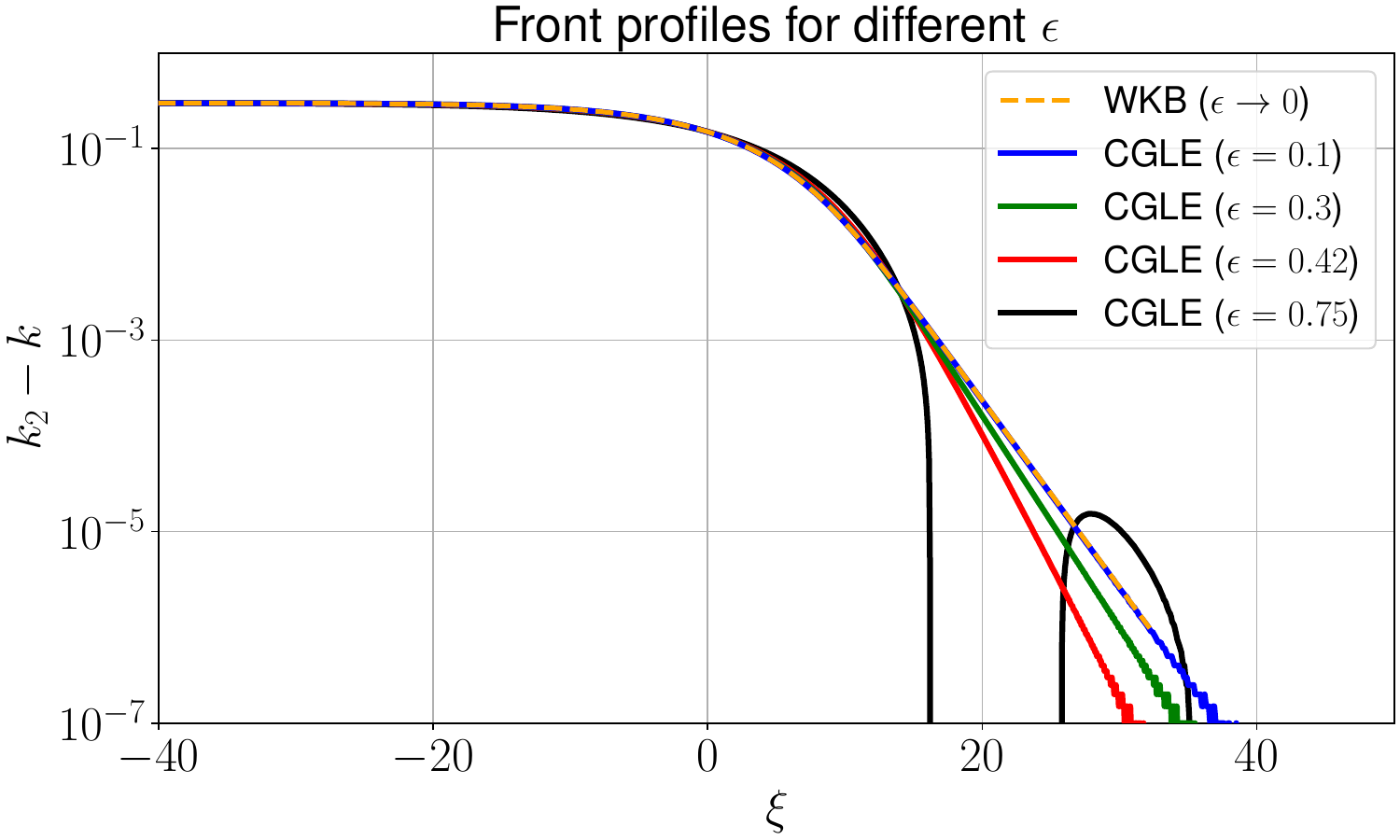}
    \caption{Comparison of front solution profiles at different values of $\epsilon$, interpolating between $k_1=0.2$ and $k_2=0.5$. For  $\epsilon=0.1$, the profile is nearly indistinguishable from the WKB limit $\epsilon \to 0$. The front increases in steepness with increasing $\epsilon$ but becomes shallower after the complex eigenvalue emerges, cf. Fig.~\ref{fig:spatial_eigenvalues_vs_eps}.} \label{fig:comparison_front_profiles_different_eps}
\end{figure}

Figure~\ref{fig:comparison_front_profiles_different_eps} illustrates the convergence of the front profile to the exact solution obtained from WKB theory as described in Sec.~\ref{sec:front_CGLE}: at $\epsilon=0.1$ and the parameter values $\alpha=0.5$, $\beta=1$, $k_1=0.2$, $k_2=0.5$ the profile is visually indistinguishable from the solution of the nonlinear diffusion equation (\ref{eq:nl_diffusion_equation_complex_2nd_order}). As $\epsilon$ increases, the front at first steepens, remaining monotonic, up to the threshold value of $\epsilon$ where complex eigenvalues emerge, beyond which the front becomes oscillatory and simultaneously shallower (see also Fig.~\ref{fig:spatial_eigenvalues_vs_eps}).

For small $\epsilon$, all eigenvalues are real, and thus collisions of chasing fronts similar to those shown previously in Fig.~\ref{fig:spacetime_plot_collision} inevitably lead to the formation of a single front, a prediction confirmed by DNS (not shown).


\subsection{Finite-time singularity in Kuramoto-Sivashinsky-like expansion }
\label{sec:relation_to_KS_eqn}
When $\alpha\ne\beta$ and the fourth-order regularization is included, the wave number equation becomes
\begin{equation}
k_\tau=\partial_\xi\left[D(k)k_\xi+(\beta-\alpha)k^2\right]
-\kappa k_{\xi\xi\xi\xi},
\end{equation}
where $D(k)=(1-3k^2)/(1-k^2)$, as before. A numerical study of this full equation is left for future work. Here we examine its relation to the Kuramoto--Sivashinsky (KS) equation \cite{kuramoto1976persistent,kuramoto1976turbulent,sivashinski1977nonlinear,michelson1977nonlinear,kuramoto1978diffusion,kuramoto1984chemical,sivashinsky1985weak}.

Expanding about a constant wave number $k_0$ via $k=k_0+v$ and retaining terms through quadratic order, the equation for $v(x,t)$ in a frame moving with velocity $c_g=2(\alpha-\beta)k_0$ reads
\begin{equation}
\begin{aligned}
v_\tau={}&D_0v_{\xi\xi}+\frac{D'_0}{2}(v^2)_{\xi\xi}\\
&+(\beta-\alpha)(v^2)_\xi-\kappa v_{\xi\xi\xi\xi},
\end{aligned}
\label{eq:KS_like}
\end{equation}
Here, $D_0=D(k_0)$ and $D'_0=-4k_0/(1-k_0^2)^2<0$ for $0<k_0<1$. This is a special case of the generalized KS equation studied by Christov and Velarde \cite{ChristovVelarde1995}, without their third-order dispersive term.

For $k_0>1/\sqrt{3}$, $D_0<0$ produces the antidiffusion characteristic of the KS equation. Unlike the standard KS equation, however, Eq.~\eqref{eq:KS_like} also contains the nonlinear diffusion term $(D'_0/2)(v^2)_{\xi\xi}$, which makes the effective diffusivity amplitude-dependent, $D_{\mathrm{eff}}(v)=D_0+D'_0v$. Since $D'_0<0$, this term enhances antidiffusion for positive perturbations while weakening it for negative ones, thereby introducing an asymmetry in their nonlinear evolution.

For $\alpha=\beta$, Eq.~\eqref{eq:KS_like} is of the form $v_\tau=\mu_{\xi\xi}$ with
\begin{equation}
\begin{aligned}
\mu&=D_0v+\frac{D'_0}{2}v^2-\kappa v_{\xi\xi}
=\frac{\delta\mathcal F}{\delta v},\\
\mathcal F[v]&=\int\left[
\frac{D_0}{2}v^2+\frac{D'_0}{6}v^3
+\frac{\kappa}{2}v_\xi^2
\right]d\xi.
\end{aligned}
\end{equation}
For periodic boundary conditions,
$d\mathcal F/d\tau=-\int\mu_\xi^2\,d\xi\le0$.
The (cubic) potential is unbounded below. If the expansion about $k_0\in (0,1)$ is continued to higher order, this property persists, since $D^{(n)}(k_0)<0$ for all
$n\geq1$, as follows from
$D(k)=3-(1-k)^{-1}-(1+k)^{-1}$.
In contrast, the full free-energy density $f(k)$ is bounded
below in $|k|<1$ [cf. Eq.~\eqref{eq:nonuniform_critical_points3}].
Thus polynomial truncations lose this constraint on the original model.

Numerical simulations of Eq.~\eqref{eq:KS_like} with small-amplitude random initial conditions and periodic boundary conditions consistently exhibit rapid amplitude growth suggestive of finite-time blow-up that persists under numerical refinement (not shown). This behavior contrasts with the spatiotemporal chaos characteristic of the standard KS equation \cite{aranson2002world}. 

Equation~\eqref{eq:KS_like} also differs from the convective Cahn--Hilliard equation \cite{emmott,golovin,watson,podolny}, whose stabilizing cubic nonlinearity $(v^3)_{\xi\xi}$ corresponds to a potential bounded below, and from the conserved KS equation \cite{frisch2006effect,gelfand2015one}, whose nonlinear term $(v_\xi^2)_{\xi\xi}$ is associated with coarsening rather than the amplitude growth observed here.

\section{Conclusions}
\label{sec:conclusions}
In this work we showed that a single scalar nonlinear wave number equation derived from a WKB reduction of the GLE organizes several seemingly disparate dynamical phenomena, including unstable stationary states, finite-time collapse associated with phase slips, and shock formation in the presence of complex coefficients.

For the RGLE, the reduced dynamics possesses a conserved gradient structure analogous to a Cahn–Hilliard system. Within this framework we constructed exact stationary wave number pulses and spatially periodic modulations, and showed that all such nonuniform states are linearly unstable. Localized pulses necessarily straddle the Eckhaus boundary, connecting stable plane waves with identical far-field wave numbers, while reaching into the Eckhaus-unstable regime at their core. In the Eckhaus-unstable regime, it is shown that for suitable initial conditions, the reduced equation develops finite-time singularities. We derived a self-similar description of this collapse within the regularized reduced equation, predicting that the local maximum $k_{\max}$ of the wave number in the approach to the phase slip satisfies
\begin{equation}
1-k_{\max}(\tau) \sim (\tau_* - \tau)^{1/2},
\end{equation}
where $\tau_*$ is the time of the phase slip, together with similarity profiles governed by a nonlinear fourth-order boundary value problem. These results are supported by our DNS of the reduced equation.

A limitation of the present analysis is that the higher-order regularization is modeled using a hyperdiffusive term with a constant prefactor. A systematic higher-order WKB reduction instead yields state-dependent coefficients that also become singular at the phase slip. Moreover, amplitude dynamics are expected to become important as the phase slip is approached, invalidating the slaving relation underlying the reduced description. The effects of these limitations on singularity formation remain to be determined.

Extending the reduction to the CGLE introduces a Burgers-type nonlinear term in the wave number dynamics. This term drives wave steepening, which supports traveling fronts that connect plane waves of arbitrary wave numbers within the Eckhaus-stable band, where the front acts as a wave sink, and within the Eckhaus-unstable band, where the front acts as a wave source. We obtained an exact expression for these fronts and demonstrated convergence to the predicted profiles in direct numerical simulations of the full CGLE in the large-scale, near real-coefficient limit. Away from this limit, the fronts develop an oscillatory structure, which we explained in terms of a transition in the dominant spatial eigenvalue from real to complex values. The resulting solutions go beyond the classical Nozaki-Bekki solutions.

We also examined a Kuramoto-Sivashinsky-like expansion of the reduced equation in the Eckhaus-unstable regime. Unlike the standard Kuramoto-Sivashinsky equation, the expansion contains a destabilizing nonlinear diffusion term and exhibits rapid amplitude growth suggestive of finite-time blow-up, which is consistent with the singularity formation observed in the reduced wave number equation.

Our results suggest potentially observable signatures in effectively one-dimensional pattern-forming systems. For example, Ref.~\cite{spina1998confined} uses direct numerical simulations to identify propagating pattern compressions in thermosolutal convection akin to our homoclinic connections in wave number. 
Moreover, phase slips are ubiquitous in all pattern-forming systems, allowing the system to respond to parameter changes and relax to the energetically favored wave number. In particular, self-similar dynamics during the approach to a phase slip could be investigated by tracking the evolution of the local pattern wavelength and the spatial extent of the region in which the amplitude decreases. Although the scaling exponents derived here may depend on the higher-order regularization employed, self-similar behavior near phase slips may persist more generally. For traveling-wave patterns, our exact front solutions predict characteristic spatial profiles, including a transition from monotonic to oscillatory exponential tails. These predictions could be tested in systems described by the CGLE.

Several directions for future work remain. The predicted self-similar approach to phase slips should be compared with the full GLE in the regime $\epsilon\ll1$. For the CGLE, numerical continuation in $\epsilon$ of the two-parameter family of exact fronts identified in Sec.~\ref{sec:front_CGLE}, together with an analysis of their existence and stability, could establish their relevance beyond the asymptotic regime and potentially identify new coherent structures, for which no exact solutions are known. Extending the rigorous approximation results of Melbourne and Schneider \cite{melbourne2004phase} to complex coefficients, and generalizing the reduction to higher spatial dimensions, where anisotropic phase diffusion and topological defects may introduce additional complications, are further open problems.

\begin{acknowledgments}
We acknowledge support from the National Science Foundation under DMS-2308337 (E.K. and Y.L.) and a US Dynamics Days travel stipend (Y.L.). We acknowledge helpful discussions with C. Liu and N. Verschueren van Rees, and thank P. Howard for helpful feedback on the spectral stability analysis of stationary states.
\end{acknowledgments}

\appendix
\section{Higher--order WKB correction for the real Ginzburg--Landau equation}
\label{app:hi_ord_wkb}
This appendix derives the $O(\epsilon^2)$ WKB correction to the nonlinear phase diffusion equation for the RGLE, yielding a fourth-order conservation law for the local wave number. Starting with the RGLE, 
\begin{equation}
\epsilon^2 u_\tau = u+\epsilon^2 u_{\xi\xi}-|u|^2u \label{eq:rgle_app}
\end{equation}
written in slow variables $\xi=\epsilon x$, $\tau=\epsilon^2 t$, we seek a modulated wavetrain solution of the form
\begin{equation}
u(\xi,\tau)=R(\xi,\tau)\exp\!\left(\frac{i}{\epsilon}\phi(\xi,\tau)\right)
\end{equation}
with local wave number $k\equiv\phi_\xi$. We write 
\begin{align}
(R,\phi) &= (R_0,\phi_0)+\epsilon^2 (R_2,\phi_2)+O(\epsilon^4),
\end{align}
while assuming $k_0=(\phi_0)_\xi=O(1)$ and  $R_0=O(1)$. We note that only $\epsilon^2$ appears in Eq.~(\ref{eq:rgle_app}) and this is therefore the natural parameter for the expansion.

The imaginary part gives the exact phase equation
\begin{equation}
\phi_\tau = 2\frac{R_\xi}{R}k+k_\xi .
\label{eq:phase_exact}
\end{equation}
Taking the real part, one finds that at leading order the amplitude remains slaved to the wave number, $R_0^2=1-k_0^2$. Furthermore, differentiating the leading-order terms in \eqref{eq:phase_exact} with respect to $\xi$ gives the familiar nonlinear diffusion equation for the wave number,  Eq.~\eqref{eq:nl_diffusion_equation_complex_2nd_order},
\begin{equation}
k_{0\tau}=\partial_\xi\!\left(D(k_0)k_{0\xi}\right),\qquad
D(k_0)=\frac{1-3k_0^2}{1-k_0^2}.
\label{eq:leading}
\end{equation}

At $O(\epsilon^2)$, the real part yields
\begin{equation}
R_2=
-\frac{R_{0\tau}-R_{0\xi\xi}}{2R_0^2}
-\frac{k_0}{R_0}k_2 .
\end{equation}
The decomposition $\phi=\phi_0+\epsilon^2\phi_2+O(\epsilon^4)$, i.e., $k=k_0+\epsilon^2k_2$, is not unique: the transformation
$\phi_0\mapsto\phi_0+\epsilon^2\chi$, $\phi_2\mapsto\phi_2-\chi$ leaves $\phi$ unchanged. 
We therefore perform a near-identity transformation to achieve
\begin{equation}
k_2=\phi_{2\xi}\equiv0,
\end{equation}
    which implies that the wave number $k=k_0+O(\epsilon^4)$, while assigning all $O(\epsilon^2)$ corrections to $R$:
\begin{equation}
R_2=
-\frac{R_{0\tau}-R_{0\xi\xi}}{2R_0^2}.
\end{equation}
Using $R_0=\sqrt{1-k_0^2}$ together with \eqref{eq:leading}, one obtains
\begin{equation}
R_2=-\frac{k_0^3}{R_0^5}\,k_{0\xi\xi}-\frac{1+3k_0^2}{2R_0^7}\,k_{0\xi}^2.
\end{equation}
Differentiating \eqref{eq:phase_exact} with respect to $\xi$ and including terms up to $O(\epsilon^2)$ now leads to 
\begin{equation}
k_{0\tau}
=
\partial_\xi\!\big(D(k_0)k_{0\xi}\big)
+
2\epsilon^2\,\!\left(
k_0 \left(\frac{R_2}{R_0}\right)_\xi
\right)_\xi
+O(\epsilon^4).
\end{equation}
Using $k=k_0+O(\epsilon^4)$, we may henceforth replace
$k_0$ by $k$ to the order considered. Eliminating $R_0$ and $R_2$ then yields the closed
wave number equation
\begin{align}
k_\tau
&=
\partial_\xi\!\big(D(k)k_\xi\big)
\nonumber\\
&\quad
-\epsilon^2\partial_\xi
\Big(
E(k)k_{\xi\xi\xi}
+F_1(k)k_\xi k_{\xi\xi}
+F_2(k)(k_\xi)^3
\Big) + O(\epsilon^4),
\end{align}
with coefficients
\begin{align}
E(k)&=\frac{2k^4}{(1-k^2)^3},\\
F_1(k)&=\frac{2k(3k^4+6k^2+1)}{(1-k^2)^4},\\
F_2(k)&=\frac{k^2(18k^2+14)}{(1-k^2)^5}.
\end{align}

Note that $E(k)>0$, indicating that the fourth-order term is indeed regularizing the dynamics. The divergence of these coefficients as $k\to\pm1$ reflects the loss of uniform validity of the WKB expansion near phase-slip events, where $R_0\to0$. We note, as is common in bifurcation theory \cite{crawford1991introduction}, that near-identity coordinate transformations can be performed to modify the nonlinear terms in the expansion, but this is not discussed further here.

\section{Proofs of certain properties of steady states}
\label{app:proofs}
Here we provide proofs of certain statements from Sec.~\ref{sec:steady_solutions}.

\begin{enumerate}
    \item \textbf{Claim}: There exists no solution of Eqs.~(\ref{eq:nonuniform_critical_points1})--(\ref{eq:nonuniform_critical_points3}) that is a smooth heteroclinic orbit connecting two wave numbers $k_1\neq k_2$ at $\xi \to \pm \infty$.\\
\textbf{Proof}: The function $k(\xi)$ can only asymptote to $k_{1,2}$ with $|k_{1,2}|<1/\sqrt{3}$. This follows from the linearization about the asymptotic wave numbers $k_{1,2}$, indicating that small deviations $\tilde{k}$ from $k_{1,2}$ are of the form $\tilde{k}\propto \exp\left[\pm \sqrt{D(k_{1,2})}\xi\right]$, which can only converge to zero if $D(k_{1,2})>0$, i.e., $|k_{1,2}|<1/\sqrt{3}$; otherwise one finds pure oscillations rather than exponential decay. Therefore, we can restrict our attention to $k_1,k_2\in (-1/\sqrt{3},1/\sqrt{3})$, where $f''(k)>0$. \\
We require $f(k_1)=f(k_2)=0$ ($k_\xi$ vanishes asymptotically) and $f'(k_1)=f'(k_2)=0$ ($k_{\xi\xi}$ vanishes asymptotically). However, since $f''(k)>0$, $f'(k)$ is strictly increasing and vanishes at most for one value of $k$ on this interval. Therefore, the only possibility is $k_1=k_2$, i.e., a homoclinic orbit. $\square$
   \item \textbf{Claim}: Nonconstant homoclinic orbits smoothly connecting given asymptotic wave numbers $k_\infty\in(-1/\sqrt{3},1/\sqrt{3})$ at $\xi\to-\infty$ and $\xi\to\infty$ have a single extremum exceeding $1/\sqrt{3}$ in magnitude.\\
\textbf{Proof}: Convergence $k\to k_\infty$ as $\xi\to\pm\infty$ requires that $f(k_\infty)=f'(k_\infty)=0$ and, by assumption, $f''(k_\infty)>0$.

Consider $k_\infty\geq 0$ and a homoclinic orbit with a positive maximum. Then, for $k\geq k_\infty$,
$f'(k)=\int_{k_\infty}^k D(s)\,ds$, where $D(s)$ changes sign exactly once between $k_\infty$ and $1$, namely at $1/\sqrt{3}$. Thus $f'(k)$ strictly increases from zero for $k\in[k_\infty,1/\sqrt{3})$ and decreases monotonically for $k\in(1/\sqrt{3},1)$. Consequently, $f'$ has at most one zero
in $(k_\infty,1)$, so $f$ first increases and then decreases if a second zero exists.

The maximum $k_0=k(\xi_0)$ of the assumed homoclinic orbit satisfies $f(k_0)=k_\xi^2/2=0$. Since $f(k)>0$ for $k\in(k_\infty,1/\sqrt{3}]$, necessarily $k_0>1/\sqrt{3}$. Moreover, the monotonicity properties of $f'$ imply that this second zero of $f$ is unique and satisfies $f'(k_0)<0$. Hence $k_{\xi\xi}(\xi_0)=f'(k_0)<0$, so the orbit $k(\xi)$ has exactly one local maximum.

The corresponding negative wave number solutions follow by the symmetry $k\to-k$.
$\square$

    \item \textbf{Claim:} Let $k_\star(\xi)$ be a smooth stationary solution of
\begin{equation}
k_\tau=\partial_{\xi\xi}\mu,\quad
\mu=f'(k)-k_{\xi\xi},\quad f''(k)=D(k) \label{eq:conserved_gradient_dynamics_appendix}
\end{equation}
satisfying
\[
k_\star(\xi)\to k_\infty \quad \text{as }|\xi|\to\infty,
\quad D(k_\infty)>0,
\]
and assume that $k_\star$ has exactly one local maximum. Then $k_\star$ is linearly unstable.

\textbf{Proof:}
Substituting
$k=k_\star(\xi)+v(\xi)e^{\lambda\tau}$
and linearizing gives
\begin{equation}
\begin{aligned}
\lambda v &= \partial_{\xi\xi}\mathcal Lv,\\
\mathcal L &=
-\partial_{\xi\xi}+D(k_\star(\xi)).
\end{aligned}
\label{eq:lin}
\end{equation}
We seek an eigenvalue $\lambda>0$.

Stationarity implies
$f'(k_\star)-k_{\star,\xi\xi}=\mathrm{const.}$
Differentiation gives the translation mode
\begin{equation}
\mathcal L k_{\star,\xi}=0.
\end{equation}
Writing $D_\infty=D(k_\infty)$, we have
\begin{equation}
\begin{aligned}
\mathcal L &=
-\partial_{\xi\xi}+D_\infty+V(\xi),\\
V(\xi)&\equiv D(k_\star(\xi))-D_\infty. \label{eq:def_potential_V}
\end{aligned}
\end{equation}
Since $V$ decays exponentially at infinity, Weyl's theorem
\cite[Theorem~14.6]{HislopSigal},
see also \cite{cycon1987schrodinger}, gives the essential spectrum
\begin{equation}
\sigma_{\mathrm{ess}}(\mathcal L)
=[D_\infty,\infty).
\end{equation}
Thus the translation eigenvalue $0$ lies below
the essential spectrum and is discrete.
Since $k_{\star,\xi}$ has exactly one zero,
the Sturm oscillation theorem
\cite[Thm.~9.40]{TeschlQMM} implies that
$0$ is not the lowest eigenvalue. Hence
there is a localized eigenfunction $\psi_0$
such that
\begin{equation}
\mathcal L\psi_0=\nu_0\psi_0,
\qquad \nu_0<0.
\end{equation}

This negative eigenvalue of $\mathcal{L}$ does not immediately imply instability, because an eigenfunction of $\mathcal L$ need not be an eigenfunction of $\partial_{\xi\xi}$.

Let $h$ be any smooth, compactly supported (and hence bounded) function with
$\int_{\mathbb R}h\,d\xi=1$. We define the following parameterized test function,
\begin{equation}
v_R(\xi)\equiv \psi_0(\xi)
-\frac{M}{R}h(\xi/R),
\end{equation}
where $M\equiv \int_{\mathbb R}\psi_0\,d\xi$, satisfying $\int_{\mathbb R}v_R\,d\xi=0$.

Letting $g_R(\xi)\equiv \frac{M}{R}h(\xi/R)$ and using
$\mathcal L\psi_0=\nu_0\psi_0$, we obtain the identity
\begin{equation}
\langle v_R,\mathcal Lv_R\rangle ={}\nu_0\|\psi_0\|_{L^2}^2 -2\nu_0\langle g_R,\psi_0\rangle +\langle g_R,\mathcal Lg_R\rangle,  
\end{equation}
where $\langle\dots\rangle$ denotes the $L^2$ inner product.
Direct estimates of the latter two terms yield
\begin{equation}
\langle v_R,\mathcal Lv_R\rangle
=\nu_0\|\psi_0\|_{L^2}^2+O(R^{-1}), \label{eq:vRLvR_estimate}\end{equation}
which becomes negative for sufficiently large $R$.

Following Howard
\cite[Lem.~3.1]{Howard2010}, we introduce
the self-adjoint operator
\begin{equation}
T\equiv B^{1/2}\mathcal LB^{1/2},
\end{equation}
in terms of $B=-\partial_{\xi\xi}$, where $B^{1/2}$ is defined by its Fourier
symbol $|q|$.

We let $w_R\equiv B^{-1/2}v_R$. Since $v_R$ has zero integral and a finite first moment, its Fourier transform $\widehat v_R(q)=O(q)$ as $q\to0$; hence $\|w_R\|_{L^2}^2=\int_{\mathbb R}|\widehat v_R(q)|^2/q^2\,dq<\infty$. The function $w_R$ satisfies 
\begin{equation}
\langle w_R,Tw_R\rangle
=\langle v_R,\mathcal Lv_R\rangle <0,
\end{equation}
where Eq.~\eqref{eq:vRLvR_estimate} was used.  

By a standard variational principle
\cite[Thm.~2.20]{TeschlQMM} the lower edge of the spectrum of $T$ satisfies
\begin{equation}
\inf\sigma(T)
\le
\frac{\langle w_R,Tw_R\rangle}
{\|w_R\|_{L^2}^2}<0.
\end{equation}

To establish that the negative spectrum of $T$
contains an eigenvalue, write
\begin{equation}
\begin{aligned}
T &= T_\infty+K,\\
T_\infty &= B^2+D_\infty B,\\
K &= B^{1/2}VB^{1/2},
\end{aligned}
\end{equation}
where $V$ is as defined in \eqref{eq:def_potential_V}.
The constant-coefficient operator $T_\infty$
has the Fourier symbol $q^4+D_\infty q^2$.
Since $D_\infty>0$, its spectrum is
$[0,\infty)$. Moreover, the smooth,
exponentially decaying potential $V$
produces a localized perturbation that
leaves the essential spectrum unchanged
by Weyl's theorem. Therefore,
\begin{equation}
\sigma_{\mathrm{ess}}(T)
=\sigma_{\mathrm{ess}}(T_\infty)
=[0,\infty).
\end{equation}

Since $\inf\sigma(T)<0$, there is an
isolated eigenvalue
\begin{equation}
Tw=-\lambda w,\qquad \lambda>0.
\end{equation}
Setting $v=B^{1/2}w$ gives us back the original linear stability problem \eqref{eq:lin},
\begin{equation}
-B\mathcal Lv = \partial_{\xi\xi}\mathcal L v=\lambda v,
\end{equation}
where we have now established the existence of an eigenvalue $\lambda >0$. Consequently, the stationary pulse is linearly unstable.
\hfill$\square$

\item \textbf{Claim:}
Let $k_\star(\xi)$ be a smooth $P$--periodic stationary solution of
Eq.~\eqref{eq:conserved_gradient_dynamics_appendix} which oscillates
between $k_1<k_2$ on each period. Then $k_\star$ is linearly unstable.

\medskip
\medskip\noindent\textbf{Proof.}
Linearizing about $k_\star$ gives
\begin{equation}
\lambda v=\partial_{\xi\xi}\mathcal Lv,\qquad
\mathcal L=-\partial_{\xi\xi}+D(k_\star(\xi)).
\label{eq:periodic_lin}
\end{equation}
Stationarity implies $\mu_{\xi\xi}=0$. Since $\mu$ is periodic, it is constant; differentiation with respect to $\xi$ yields
\begin{equation}
\mathcal L k_{\star,\xi}=0.
\end{equation}
Consider $\mathcal L$ on $[0,P]$ with periodic boundary conditions. Its lowest periodic eigenvalue is simple, with an eigenfunction of fixed sign (\cite[Thm.~6.9.1]{Eastham1973}). Since $k_{\star,\xi}$ changes sign, it cannot be the ground state. Consequently, $\mathcal L$ has a negative periodic eigenvalue with a real eigenfunction,
\begin{equation}
\mathcal L\psi_0=\nu_0\psi_0,
\qquad \nu_0<0.
\end{equation}

To assess stability on $\mathbb R$, we use Floquet--Bloch theory \cite{kuchment1993floquet} and write $v=e^{i\theta\xi}v_\theta$, where $v_\theta(\xi+P)=v_\theta(\xi)$ and $\theta\in[-\pi/P,\pi/P]$. The spectrum on $\mathbb R$ is the union of the spectra of these Bloch problems. The eigenvalue equation is
\begin{equation}
\begin{aligned}
\lambda v_\theta&=-B_\theta\mathcal L_\theta v_\theta,\\
B_\theta&=-(\partial_\xi+i\theta)^2,\\
\mathcal L_\theta&=-(\partial_\xi+i\theta)^2+D(k_\star). \label{eq:Bloch_problem}
\end{aligned}
\end{equation}
Here and below, $\langle\cdot,\cdot\rangle$ denotes the $L^2(0,P)$ inner product. Using the real periodic eigenfunction $\psi_0$ (of $\mathcal L$) as a trial function for $\mathcal L_\theta$ gives
\begin{equation}
\begin{aligned}
\langle\psi_0,\mathcal L_\theta\psi_0\rangle
&=\langle\psi_0,\mathcal L\psi_0\rangle
 +\theta^2\|\psi_0\|_{L^2(0,P)}^2\\
&=(\nu_0+\theta^2)\|\psi_0\|_{L^2(0,P)}^2,
\end{aligned}
\end{equation}
where the term linear in $\theta$ vanishes by periodicity. Hence, for sufficiently small nonzero $|\theta|$,

\begin{equation}
\langle\psi_0,\mathcal L_\theta\psi_0\rangle<0.
\end{equation}

For $0<|\theta|<\pi/P$, $B_\theta$ is positive and invertible. As in the proof of linear instability for the homoclinic pulse, we introduce the self-adjoint operator
\begin{equation}
T_\theta=B_\theta^{1/2}\mathcal L_\theta B_\theta^{1/2}.
\end{equation}
Setting $w_\theta=B_\theta^{-1/2}\psi_0$, one finds
\begin{equation}
\langle w_\theta,T_\theta w_\theta\rangle
=\langle\psi_0,\mathcal L_\theta\psi_0\rangle<0.
\end{equation}
By the same standard variational principle used in previous proof \cite[Thm.~2.20]{TeschlQMM}, the lower edge of the spectrum of $T_\theta$ obeys
\begin{equation}
\inf\sigma(T_\theta)
\le\frac{\langle w_\theta,T_\theta w_\theta\rangle}{\|w_\theta\|_{L^2(0,P)}^2}<0.
\end{equation}
Since \(T_\theta\) has discrete eigenvalues on the finite periodic cell, its lowest eigenvalue is \(-\lambda=\inf\sigma(T_\theta)<0\), with \(\lambda>0\), i.e., $T_\theta w=-\lambda w$. Setting $v_\theta=B_\theta^{1/2}w$ yields Eq.~\eqref{eq:Bloch_problem}. Thus a positive growth rate is present in the Floquet--Bloch spectrum on $\mathbb R$, proving linear instability.\hfill$\square$
\end{enumerate}

\section{Exact equation for the front profile}
\label{app:exact_front}
Here we provide an analytical expression for the integral in Eq.~(\ref{eq:front_profile_integral_z_of_k}) describing the shape of the front between $k_1$ and $k_2$ in the case of complex coefficients. First, we rewrite the integrand in partial fraction form
\begin{align}
&\frac{1-3k^{2}}{(1-k^{2})(A-c\,k-(\beta-\alpha)k^{2})}
\nonumber \\&=\frac{a}{1-k}+\frac{b}{1+k}
 +\frac{d}{k-k_{1}}
 +\frac{f}{k-k_{2}},
\end{align}
where 
\begin{align}
a &= \frac{1}{(\beta-\alpha)(1-k_{1})(1-k_{2})}, \nonumber\\
b &= \frac{1}{(\beta-\alpha)(1+k_{1})(1+k_{2})},\nonumber \\
d &=
-\frac{1-3k_{1}^{2}}{(\beta-\alpha)(1-k_{1}^{2})(k_{1}-k_{2})}, \nonumber\\f &=
-\frac{1-3k_{2}^{2}}{(\beta-\alpha)(1-k_{2}^{2})(k_{2}-k_{1})}.
\end{align}
This leads to the result
\begin{equation}
z(k) - z_0
= -a\ln|1-k|
  + b\ln|1+k|
  + d\ln|k-k_{1}|
  + f\ln|k-k_{2}|,
\end{equation}
where $z_{0}$ is an arbitrary constant of integration. For $\beta>\alpha$, $z(k)$ is a monotonically increasing function of $k$, while for $\beta<\alpha$, it is monotonically decreasing. Therefore, the transcendental relation can easily be inverted numerically to obtain $k(z)$. The result is the expression shown in Fig.~\ref{fig:front_second_order} along with the numerically evaluated integral.

\bibliographystyle{unsrt}
\bibliography{references}

\end{document}